\documentclass[sn-mathphys-num]{sn-jnl}

\usepackage{graphicx}%
\usepackage{multirow}%
\usepackage{amsmath,amssymb,amsfonts}%
\usepackage{amsthm}%
\usepackage{mathrsfs}%
\usepackage[title]{appendix}%
\usepackage{xcolor}%
\usepackage{textcomp}%
\usepackage{manyfoot}%
\usepackage{booktabs}%
\usepackage{algorithm}%
\usepackage{algorithmicx}%
\usepackage{algpseudocode}%
\usepackage{listings}%
\usepackage[skip=0pt]{caption}
\usepackage{float}
\usepackage{romannum}
\usepackage{aas_macros}

\def\be{\begin{equation}}
\def\ee{\end{equation}}
\def\kms{{\rm \,km\,s^{-1}}}

\def\s{{\rm \,s}}
\def\Gyr{{\rm \,Gyr}}

\def\cm{{\rm \,cm}}

\def\kpc{{\rm \,kpc}}
\def\arcmin{{\rm \,arcmin}}
\def\erg{{\rm \,erg}}
\def\eV{{\rm \,eV}}
\def\keV{{\rm \,keV}}

\newcommand{\dd}{{\rm d}}

\theoremstyle{thmstyleone}%
\theoremstyle{thmstyletwo}%

\theoremstyle{thmstylethree}%

\begin{document}

\graphicspath{{figures/}}

\title[Gas Kinematic Mapping of the Perseus Cluster]{Disentangling Multiple Gas Kinematic Drivers in the Perseus Galaxy Cluster}


\author[1]{\fnm{The XRISM} \sur{collaboration}}

\affil[1]{The list can be found on the last page}


\abstract{
Galaxy clusters, the Universe’s largest halo structures, are filled with 10-100 million degree X-ray-emitting gas. Their evolution is shaped by energetic processes such as feedback from supermassive black holes (SMBHs) and mergers with other cosmic structures \cite{Vikhlinin2014,Fabian2012,Kravtsov2012}. The imprints of these processes on gas kinematic properties remain largely unknown, restricting our understanding of gas thermodynamics and energy conversion within clusters \cite{Simionescu2019}. High-resolution spectral mapping across a broad spatial-scale range provides a promising solution to this challenge \cite{Hitomim2016Nature,Hitomi2018U}, enabled by the recent launch of the XRISM X-ray Observatory \cite{Tashiro2020}.
Here, we present the kinematic measurements of the X-ray-brightest Perseus cluster with XRISM, radially covering the extent of its cool core.
We find direct evidence for the presence of at least two dominant drivers of gas motions operating on distinct physical scales: a small-scale driver in the inner $\sim60$ kpc, likely associated with the SMBH feedback; and a large-scale driver in the outer core, powered by mergers.
The inner driver sustains a heating rate at least an order of magnitude higher than the outer one. This finding suggests that, during the active phase, the SMBH feedback generates turbulence, which, if fully dissipated into heat, could play a significant role in offsetting radiative cooling losses in the Perseus core.
Our study underscores the necessity of kinematic mapping observations of extended sources for robust conclusions on the properties of the velocity field and their role in the assembly and evolution of massive halos. It further offers a kinematic diagnostic for theoretical models of SMBH feedback.
}


\maketitle

The Perseus cluster, located $\simeq 78$ Mpc away, is the X-ray brightest nearby cluster of galaxies.
It has been extensively studied with multi-wavelength observations, including from all major X-ray telescopes, over the past few decades \cite[e.g.,][]{Forman1972,Conselice2001,Fabian2003,Churazov2003, Urban2014,Canning2014,Hitomim2016Nature,Sanders2020,vanWeeren2024}, and is considered to be a textbook example of radio-mode Active Galactic Nucleus (AGN) feedback \cite[e.g.,][]{Boehringer1993,Churazov2000,Fabian2012}.
The SMBH residing in the cluster's Brightest Cluster Galaxy (BCG), NGC 1275, interacts with the surrounding gas through the inflation of multiple bubbles of relativistic plasma, which appear as cavities in the X-ray image (Extended Data Fig.~\ref{fig:binning2}a). The initial rapid inflation of the bubbles generates weak shocks and sound waves \cite{Fabian2003}. As the bubbles rise buoyantly in the cluster's atmosphere, they displace and mix the hot gas, driving gravity waves and turbulence \cite{Zhuravleva2014,Zhang2018,Zhang2022}. The global, modest-degree east-west asymmetry of the gas indicates an ongoing merger with infalling galaxies and groups of galaxies \cite{Churazov2003,Simionescu2012,Walker2018}, producing sloshing motions of the gas \cite{ZuHone2011,Ichinohe2019,Bellomi2024} and creating spiral structures in the cluster's residual image (Extended Data Fig.~\ref{fig:binning2}a).

All these processes deposit kinetic energy into the gas, a portion of which eventually dissipates into heat directly or through intermediate processes. This energy conversion is particularly important in the core regions of many galaxy clusters, including Perseus, where radiative energy losses occur on short timescales. Without sources of heat, this would lead to vigorous star formation and accumulation of orders of magnitude more cold gas than is observed (\cite{Lea1976,Cowie1977,Fabian1977,Peterson2006}; although see \cite{Fabian2022} for possibilities of more cooling than previously thought). To determine the dissipation rate of gas motions, which can then be directly compared to cooling losses, one needs to measure both velocity amplitudes, $v_{\rm k}$, and their corresponding length scales, $\ell_{\rm k}=1/k$, where $k$ is a wavenumber. The density-normalized heating rate in the gas is then $\propto v_{\rm k}^3/\ell_{\rm k}$ under the common assumption of motions being injected on large scales and cascading down to smaller scales. For a particular case of Kolmogorov-type turbulence, $v_{\rm k}^3/\ell_{\rm k}$ remains the same at all scales within the inertial range. In practice, however, this estimation is non-trivial, partly due to the potential presence of multiple energy-injection processes in the intracluster medium (ICM).

\begin{figure*}
\centering
\includegraphics[width=0.9\linewidth]{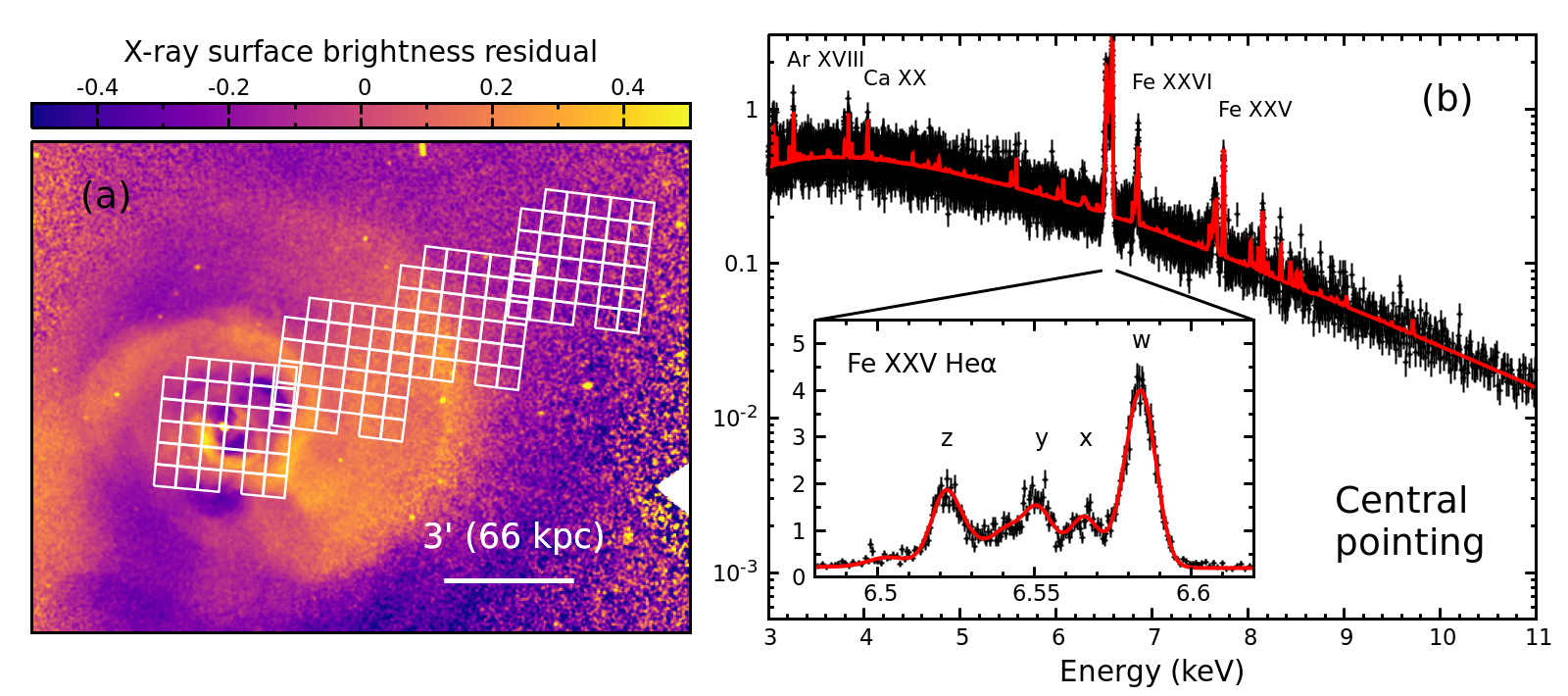}
\caption{\textbf{High-resolution X-ray image and spectrum of the Perseus cluster.} (a) X-ray surface brightness divided by the best-fit model of the mean surface brightness profile, centered at RA=49.9508 and Dec=41.5115 (J2000). The $2-8$ keV image is obtained from Chandra observations with a total cleaned exposure time $\simeq 1.4$ Ms and smoothed with a 1" Gaussian. The redshift is taken to be $z=0.017284$ (redshift of the BCG), so that 1' corresponds to a physical scale of $22\kpc$ (for a standard cosmology with $h=0.67$, $\Omega_{\rm m}=0.32$, $\Omega_{\Lambda}=0.68)$. White squares indicate XRISM/Resolve pixels for the four pointings centered at distances of $\sim 0,\ 68,\ 133$, and $200$ kpc from the center, with pixels 12 and 27 removed.  (b) XRISM/Resolve spectrum in units of $\rm counts\s^{-1}\keV^{-1}$, extracted from the entire central pointing in the $3-11\keV$ energy range, highlighting several prominent emission lines. The spectrum is based on the combined data from ObsIDs 000154000 and 000155000, resulting in a total exposure time of 98 ks. The inset provides a closer look at the strongest Fe XXV He-$\alpha$ line complex. The observed spectrum is shown in black, with the best-fit spectral model overlaid in red (Methods).}
\label{fig:perseus_map}
\end{figure*}

Hitomi observations of the Perseus cluster detected gas motions with velocity dispersion $\simeq 80-240\kms$ in the inner $\sim 60$ kpc region \cite{Hitomim2016Nature,Hitomi2018U}. Since Hitomi's observations were conducted during the commissioning phase of the mission, the spectral analysis was subject to relatively large calibration uncertainties. Here, we present the first kinematic mapping observations of the Perseus core extending up to a radius of $\simeq 250$ kpc using the X-ray microcalorimeter Resolve \cite{Ishisaki2022} onboard the X-Ray Imaging and Spectroscopy Mission (XRISM) satellite \cite{Tashiro2020}. Five observations (four pointings) along the northwestern direction were conducted in January 2024 (Fig.~\ref{fig:perseus_map}a, Methods), with the central pointing overlapping Hitomi's regions. Resolve's high energy resolution of $\sim 4.5$ eV (FWHM) allowed the detection of many individual emission lines in spectra from each region (Fig.~\ref{fig:perseus_map}b, Extended Data Fig.~\ref{fig:specs_b2L}). We divided these pointings into six sub-field of view (FOV) regions (sub-regions or regions hereafter; see Extended Data Fig.~\ref{fig:binning2}b), extracted spectra from each, and modeled them with a collisionally-ionized plasma emission model. We also included a power-law component and fluorescent lines from neutral Fe for the central AGN, non-X-ray background, and accounted for resonant scattering and scattering of photons between the sub-regions due to the broad point spread function (PSF) of the instrument. The details of the modeling are described in Methods, and the spectral fitting results are presented in Extended Data Table~\ref{tab:parameters} and Extended Data Fig.~\ref{fig:specs_b2L}.

\begin{figure*}
\centering
\includegraphics[width=0.6\linewidth]{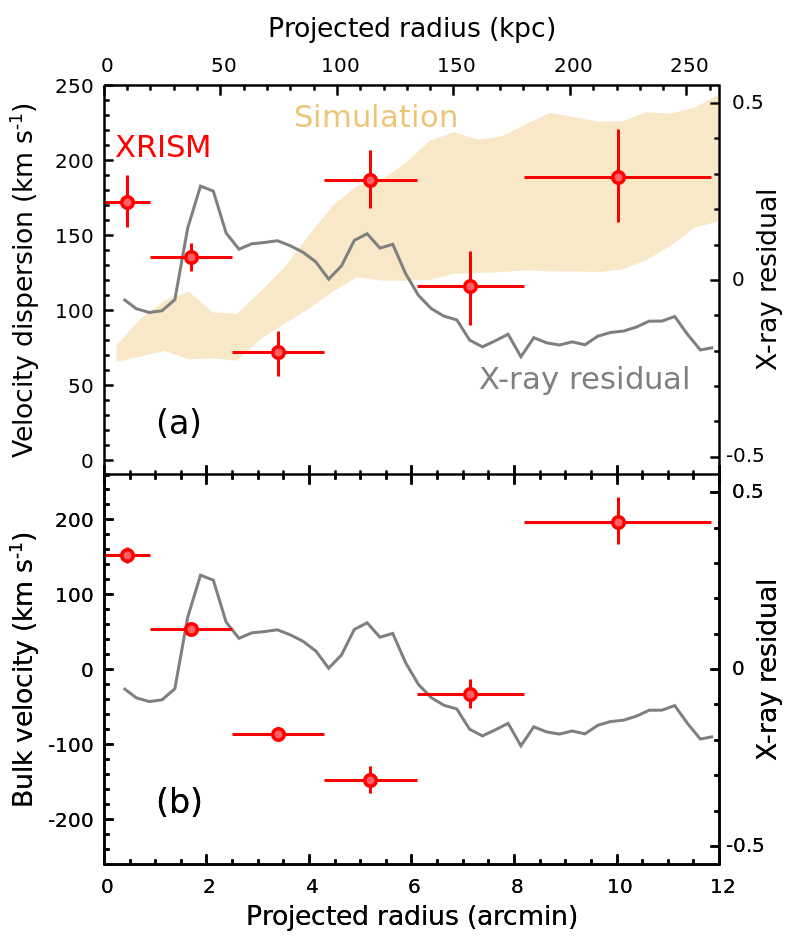}
\caption{\textbf{Radial profile of gas kinematic properties in the Perseus cluster measured with XRISM/Resolve.}
(a) LOS velocity dispersion $\sigma_{\rm ICM}$ measured from the broadening of emission lines present in the 3--11 keV energy range (red). In the innermost two regions, the width of the strongest Fe XXV He-$\alpha$ is decoupled from the broadening of other lines since its shape is affected by resonant scattering (Methods). Yellow shaded region indicates predictions from numerical simulations of stratified turbulence driven on $500$ kpc scale in the Perseus-like atmosphere (Methods), with the scatter corresponding to the standard deviation of the velocity dispersion.
(b) LOS bulk velocity of the gas obtained from centroid shifts of all emission lines in the 3--11 keV energy range relative to the heliocentrically-corrected bulk velocity of the central galaxy NGC 1275.
Vertical error bars show $1\sigma$ statistical uncertainty. Systematic uncertainties, discussed in Methods, could result in velocity variations within the $1 \sigma$ range. Calibration uncertainties of the velocity dispersion and bulk velocity are $<10\kms$ and $\simeq18\kms$ (Methods), respectively. The gray curve represents the profile of residual X-ray surface brightness in the band $0.5-8\keV$ extracted from the region covered by the XRISM pointings.}
\label{fig:vel_profs}
\end{figure*}

Fig.~\ref{fig:vel_profs} presents the measured radial profiles of the line-of-sight (LOS) velocity dispersion $\sigma_{\rm ICM}$ and bulk velocity $v_{\rm bulk}$, which are the primary focus of this study.
To obtain the bulk velocities, we applied the heliocentric correction, which was $\simeq -25\kms$ on average for each observation, and calculated the velocity relative to the heliocentric-corrected redshift of stars in the central BCG, NGC1275 ($z=0.017284$; see \cite{Hitomi2018U}).
Several trends are immediately noticeable.
(1) There is an offset of $152\pm10\kms$ between the bulk velocity of the ICM in the central sub-region and the BCG, indicating that the cluster is not completely relaxed as it was previously assumed to be \cite{Mantz2015}.
Interestingly, a comparable offset of $130\pm1\kms$ was also observed between the cold H$\alpha$ gas and the BCG in the innermost region \cite{Gendron-Marsolais2018,Vigneron2024}.
(2) The velocity dispersion profile shows two distinct drops by $\sim 60-70 \kms$ at $\simeq 50$ and $\simeq 140$ kpc.
These jumps approximately coincide with the two cold fronts visible in the cluster residual image as bright spiral structures (see also gray curves in Fig.~\ref{fig:vel_profs}), most likely associated with merger-driven sloshing of the gas.
They may reflect overlapping bulk shear motions and turbulence near the interfaces resulting from Kelvin-Helmholtz instability.
This picture is additionally supported by the bulk velocity pattern, which varies significantly across these spirals - decreasing by over $100 \kms$ across the inner front and increasing by $\sim 100 \kms$ across the outer front.
These changes in bulk velocity associated with spiral cold fronts indicate a non-zero inclination angle between the sloshing plane and the plane of the sky \cite{ZuHone2016}.
(3) Within $\sim 60$ kpc, the LOS velocity dispersion gradually increases toward the center, reaching the value of $\sim 160-200 \kms$ in the region closest to the central SMBH. The minimum velocity dispersion, $\simeq 70 \kms$, occurs at a distance of $\sim 70$ kpc from the center. From there, the dispersion gradually rises by a factor of $\sim 2$ as one moves outward, reaching levels similar to those in the innermost region at $\simeq 220$ kpc. As expected, coarser radial binning smooths the velocity gradients and jumps (Extended Data Fig.~\ref{fig:vel_profs_check}a), while still following the overall radial trends.

\begin{figure*}
\centering
\includegraphics[width=0.95\linewidth]{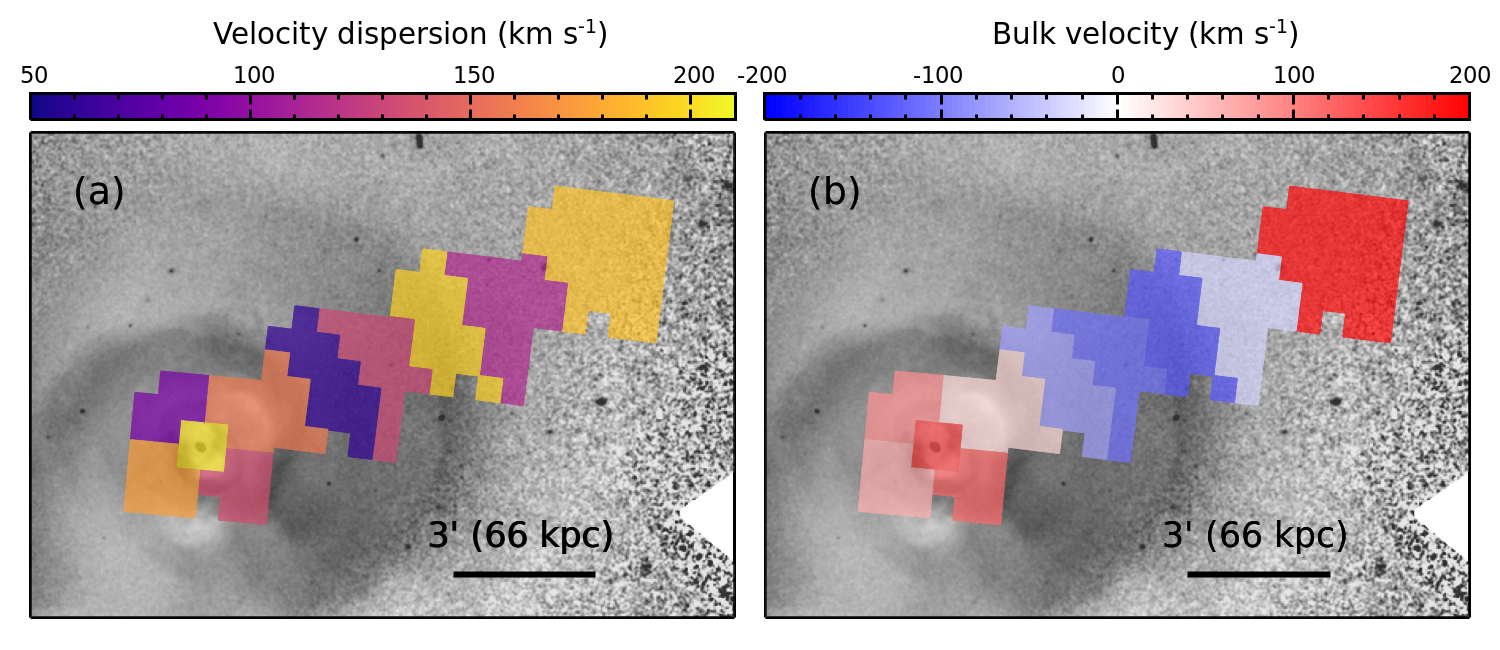}
\caption{\textbf{Kinematic maps of the hot gas in the Perseus cluster.} Velocity dispersion (a) and bulk velocity (b) maps are overlaid on the X-ray residual image (grey). The maps show azimuthal variations in the inner $\simeq30$ kpc region and radial variations in narrow bins outside this region. Statistical uncertainties for each region are shown in Extended Data Fig.~\ref{fig:map_bins_error}.}
\label{fig:map_bins}
\end{figure*}

In addition to radial profiles, we present kinematic maps of the velocity structure in finer bins, along with azimuthal variations within the inner $\sim 30$ kpc region (Fig.~\ref{fig:map_bins}, Extended Data Fig.~\ref{fig:map_bins_error}), further supporting the observed trends. These maps show small azimuthal variations, $\simeq 50 \kms$, in both velocity dispersion and bulk velocity.
The measured kinematic properties translate into 3D Mach numbers of $\simeq 0.3$ in the innermost region, $\simeq 0.1$ in the region with the lowest velocity dispersion, and intermediate values in other regions (Methods).
The corresponding non-thermal pressure fraction, namely the kinetic to total pressure ratio, varies between $\sim5\%$ and $1\%$ across the core (Extended Data Table~\ref{tab:parameters}). In the innermost $\sim 60$ kpc, the non-thermal pressure fraction is consistent with the results of Hitomi \cite{Hitomi2018U}.

Due to the sharply peaked X-ray emissivity towards the cluster center (Extended Data Fig.~\ref{fig:prof_leff}b), the physical interpretation of the X-ray radial velocity profiles must take into account spatial scales contributing to the measurements.
The LOS emissivity distribution plays a role as a velocity high-pass filter \cite{Zhuravleva2012}.
The LOS size of the region that contributes the most to the X-ray emission, or effective length scale $l_{\rm eff}$, reflects the largest measurable velocity scale in such narrow radial bins, which increases with projected distance as schematically shown in Extended Data Fig.~\ref{fig:prof_leff}a, ranging from $\sim50\kpc$ to $\sim 250$ kpc in Perseus over the XRISM radial coverage (Methods, Extended Data Fig.~\ref{fig:prof_leff}b).
This implies that the similar observed velocity dispersion of $\sim 150-200\kms$ is associated with length scales as large as $l_{\rm eff}\sim 30-60\kpc$ in the central regions, while up to $l_{\rm eff}\sim 250\kpc$ at a projected distance of $\sim 200-250$ kpc, i.e., different by a factor of $\sim5$.
Combined with the observed velocity gradient towards the outer region, these results cannot be explained by a single power-law-type power spectrum of velocities with dominant energy on large scales.

This argument is further supported by numerical simulations of stratified turbulence evolving in a static gravitational potential tailored for the Perseus cluster (Methods). In our toy model, turbulence is powered by a continuous driving force with a single injection scale. The injection scale is chosen to be rather large, between 100 and 500 kpc, to mimic a situation of merger-driven gas motions, with the injection scale comparable to the length scale of a merging object, typically hundreds of kpc \cite[e.g.,][]{Miniati2015,Shi2018}. Due to the stratification, elongated velocity structures were formed along the azimuthal direction (Extended Data Fig.~\ref{fig:sim_maps}a). We additionally considered a smaller injection scale of $50\kpc$. For each case, we calculated the projected radial profile of emissivity-weighted velocity dispersion. Fig.~\ref{fig:vel_profs} shows such a profile for the injection scale of $500\kpc$ (see Extended Data Fig.~\ref{fig:sim_maps}c for other injection scales), which reproduces the observed velocity gradient outside $\sim 70$ kpc but is significantly lower in the inner region due to the LOS emissivity weighting (e.g., $\gtrsim4\sigma$ tension with the innermost data point if taking into account both observational statistical error and azimuthal variations in simulations). Decreasing the input injection scale in simulations gives a qualitatively similar picture, until it becomes sufficiently small, i.e., $\lesssim 50$ kpc (minimum $\ell_{\rm eff}$ in Perseus), in which case the profile of velocity dispersion shows no gradients and is in tension with the observations (Extended Data Fig.~\ref{fig:sim_maps}c). Therefore, the observed enhanced turbulence in the innermost two regions implies that at least two dominant sources driving gas motions are required to explain the observed profile of velocity dispersion in the stratified Perseus atmosphere: one driving motions on large scales in the bulk of the gas, and another on smaller scales in the innermost region, forming a ``double'' velocity cascade. According to our simulations, the lowest measured velocity dispersion in the radial bin at $\simeq60-90\kpc$ provides constraints on the maximum contribution of the large-scale-driven (e.g., mergers) motions to the total kinetic energy in the innermost region, which is $\lesssim 20\%$.

\begin{figure}
\centering
\includegraphics[width=0.6\linewidth]{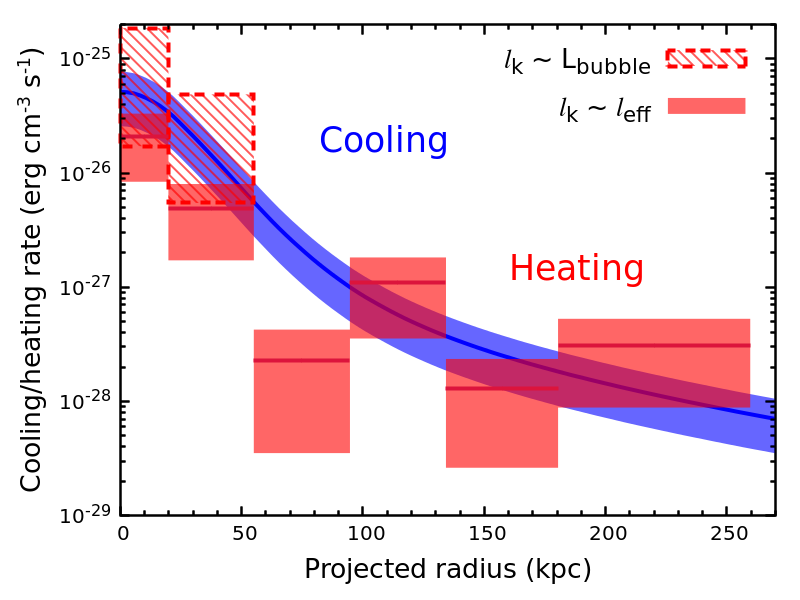}
\caption{\textbf{Estimated turbulent dissipation heating and radiative cooling rates in the Perseus cluster, assuming that the observed velocity broadening is due to turbulence.} The uncertainties on the heating rate based on $\ell_{\rm eff}$ (solid red regions) incorporate both the $1\sigma$ velocity statistical errors and effective length variations within each annulus (Extended Data Fig.~\ref{fig:prof_leff}b). Uncertainties in the heating rate derived from bubble sizes (hatched red regions) account for the same statistical errors as well as variations in bubble sizes ($L_{\rm bubble}\simeq5-30\kpc$) estimated from X-ray and radio images of the cluster. The cooling rate uncertainties (blue regions), estimated at $\pm50\%$, reflect density variations within the considered regions, based on the residual X-ray image.}
\label{fig:heating_cooling_ratio}
\end{figure}

Using directly measured velocities at each radius, we estimated the dissipation heating rate of the injected kinetic energy under the assumption of low viscosity and random motions cascading from large to small scales as $Q_{\rm heat}=C_0\rho v_{\rm k}^3/\ell_{\rm k}$, where $C_0$ is a constant coefficient (Methods), $\rho$ is the gas mass density, $v_{\rm k}$ is the velocity amplitude at a length scale $\ell_{\rm k}=1/k$. We assume for this exercise that the observed velocity dispersion $\sigma_{\rm ICM} = v_k$, and that it is dominated by turbulent motions. However, caution is needed in choosing the length scales $\ell_{\rm k}$ associated with $\sigma_{\rm ICM}$. In the outermost four regions (regions 3-6), where mergers dominate the gas kinematics, $\ell_{\rm k}$ is reasonably approximated by $l_{\rm eff}$, as discussed above, supported by the observed velocity dispersion gradient (Extended Data Fig.~\ref{fig:sim_maps}c) and mergers-driven injection scales predicted in cosmological simulations \cite{Vazza2009,Miniati2015,Shi2018}. In the inner two regions, the situation can be different, as gas kinematics is strongly influenced by AGN feedback \cite{Fabian2003,Heinrich2021}, e.g., more than $80\%$ of kinetic energy in the inner $\lesssim 20$ kpc is not associated with mergers on large scales (see Fig.~\ref{fig:vel_profs}). The scale associated with $\sigma_{\rm ICM}$ could be comparable to the size of resolved bubbles in Perseus \cite{Li2020,Zhang2022} if powered largely by radio-mode AGN feedback. This scale is lower than the effective length by a factor of $\sim 2-3$. For these inner regions, we thus conservatively estimated $Q_{\rm heat}$ using both $l_{\rm eff}$ and typical bubble sizes ($5-30$ kpc in diameter) from the X-ray and radio images of Perseus \cite{Timmerman2022}. Fig.~\ref{fig:heating_cooling_ratio} shows $Q_{\rm heat}$ for all regions, using $l_{\rm_{eff}}$ (red solid) and estimated bubble sizes (red hatched, only in regions 1-2) as $\ell_{\rm k}$. It is immediately apparent that the heating rate is not the same at all radii, which is inconsistent with a single-driven Kolmogorov-like turbulent cascade that predicts a constant heating rate within the inertial range.
Beyond region 2, it is approximately constant within $\sim 2\sigma$ with excess around 100–130 kpc linked to the sloshing cold front. In the inner two regions, it is boosted by more than an order of magnitude. This is further supporting the conclusion on multiple drivers of gas motions within the cluster core.

We can additionally compare the heating rate to the radiative cooling rate, $Q_{\rm cool}=A_0\rho^2\Lambda(T)$, where $A_0$ is a constant coefficient and $\Lambda(T)$ is the normalized cooling function (Methods), shown in Fig.~\ref{fig:heating_cooling_ratio} as a blue shaded strip. Within a factor of a few uncertainties and assuming that all motions dissipate into heat, the estimated heating rate is comparable to the cooling rate in most regions, suggesting a non-negligible role of motions in regulating thermal balance and star formation. The characteristic gas motion timescale $\ell_{\rm k}/v_{\rm k}$ is $\sim0.1$ and $\sim1\Gyr$ in the inner and outer regions, respectively. These timescales are a few times shorter than their corresponding cooling timescales, which guarantee the bubble- and merger-driven velocity fields cascade fast enough to compensate cooling losses. Region 3, a transitional region between the two drivers, is clearly underheated by turbulent dissipation even with the $3\sigma$ statistical uncertainty on $\sigma_{\rm ICM}$. Since no prominent H$\alpha$ emission has been detected in this region \cite{Gendron-Marsolais2018}, this may indicate that we have underestimated the heating associated with turbulent dissipation, or that other sources of heating may be locally operating (e.g., cosmic rays, conduction, etc.). It should also be noted that perfect cooling-heating balance, either spatially or temporally, is not required for radio-mode AGN feedback to prevent the ICM from catastrophic overcooling over the cooling timescale.

We emphasize that the heating rate estimation depends on several assumptions, including that the measured velocity dispersion primarily reflects random gas motions. This is not an unreasonable assumption given the measured levels of resonant scattering sensitive to small-scale motions \cite{Hitomi_res_scat,Zhuravleva2011}, the cluster's relatively relaxed state allowing sufficient time for post-mergers turbulence to develop \cite{Kang2024}, and indications of suppressed effective viscosity both locally and in the bulk ICM \cite{Ichinohe2019,Heinrich2024}. However, if coherent bulk motions significantly contributed to the measured $\sigma_{\rm ICM}$ \cite{ZuHone2016}, our turbulence heating rate can be overestimated. Current measurements cannot rule out small-scale mergers contributing to gas kinematics in the central regions of Perseus alongside feedback, and tailored simulations suggest this would require a finely-tuned merger configuration and precise timing \cite[e.g.,][]{ZuHone2016,Bellomi2024}. It is also important to mention that our measurements outside the inner $\sim 60$ kpc are done along a single direction direction (regions 3-6), which may not be representative of the whole ICM volume \cite{Ota2018}. The uncertainty in our knowledge of $Q_{\rm heat}$ implied in Fig.~\ref{fig:heating_cooling_ratio} also raises other important questions about the heating/cooling balance in the inner $\sim60$ kpc with regard to how quickly the kinetic energy is distributed, the resolution of which is beyond the scope of this work. In any case, the preliminary interpretations presented (prompted by our unique measurements) will need to be tested with detailed simulations and additional XRISM observations of the Perseus core region.

Regardless of the uncertainties, our main conclusion - that at least two sources on very different scales drive gas motions within the Perseus core - remains robust, thanks to the XRISM’s radial mapping observations with high spectral resolution. This finding supports the presence of AGN-powered turbulence in Perseus and offers a useful diagnostic for various AGN heating mechanisms, including turbulence and sound waves (see Methods), and sub-grid prescription for AGN feedback in numerical modeling. This is in contrast to the Centaurus cluster, where sloshing is the main driver of gas motion even in the center \cite{XRISM_Centaurus}. Meanwhile, caution is needed when characterizing volume-averaged statistical properties of velocity fields in galaxy clusters, as multiple drivers of motions may dominate at different radii. This work demonstrates that kinematic mapping studies are necessary for revealing the complexity of the velocity field in the ICM, which is important for understanding the assembly and evolution of massive halos.

\clearpage

\clearpage


\section*{Methods}

\textbf{Observations and data reduction}\\
The Perseus cluster was observed five times with XRISM/Resolve between January 21 and 26, 2024.
Two observations (obsIDs 000154000 and 000155000) were pointed at the cluster center (RA=49.9507, DEC=41.5117), while the other three (000156000, 000157000, 000158000) covered the cluster radial range $\simeq30-250\kpc$ along the NW direction (Fig.~\ref{fig:perseus_map}).
The Resolve data was reprocessed using pre-release Build 8 XRISM software, calibrated with the latest CalDB (version 10).
Default screenings of the data were applied, following the XRISM Quick Start Guide Version 2.1 \cite[see][for details]{XRISM_N123D,XRISM_NGC4151}.
After filtering, the net exposure times for obsIDs 000154000 -- 000158000 were 48, 50, 57, 93, and 131 ks, respectively.
The analysis was confined to the highest-resolution primary events (``Hp'' events), providing spectral resolution of $\simeq 4.5$ eV FWHM.

Spectra were extracted from sub-FOV detector regions described below, excluding pixel 27, which has shown unexpected gain drifts \cite{Porter2024}.
For each detector region, we created standard exposure maps, redistribution matrix files (RMFs), and auxiliary response files (ARFs, ``effective area'') by using the \texttt{xaexpmap}, \texttt{rslmkrmf}, and \texttt{xaarfgen} tools, respectively.
Throughout the analysis, we used large (``L'') RMFs, which provided consistent spectral modeling results with the extra-large (``X'') RMFs in the energy range (3-11 keV) we consider. To construct ARFs, we created an exposure-corrected 2-8 keV Chandra \texttt{ACIS} image of the Perseus cluster.
The central bright AGN was masked out from the image with the gap filled by pixel values sampled from the adjacent region.
For each detector region, multiple ARFs were generated to account for scattering into and out of the detector array footprint (spatial-spectral mixing, or SSM) due to Resolve's PSF.
Additionally, we generated point-source-model ARFs to account for the central AGN source in spectral modeling.
\\

\noindent\textbf{Spatial binning and SSM}\\
To capture the structure of the ICM velocity field, we split the pointings into multiple annular regions.
The choice of these sub-FOV regions ensured: (i) the size of each was larger than the XRISM/Resolve angular resolution (half power diameter $\simeq1.3\arcmin$); (ii) each region had at least 9000 photon counts in the $2-10\,\keV$ energy band, such that the velocity dispersions had $\lesssim30\kms$ $1\sigma$ statistical uncertainties; (iii) regions minimized mixing between bright features associated with various physical processes (Extended Data Fig.~\ref{fig:binning2}a).
The colored pixels in Extended Data Fig.~\ref{fig:binning2}b show the radial binning scheme adopted in this paper. The two innermost regions (regions 1-2) are dominated by bubbles and other structures likely produced by AGN feedback, while the boundaries between regions 2/3 and 4/5 align with the two prominent sloshing cold fronts.

Resolve spectra were extracted from each sub-FOV detector region.
However, due to the mirror PSF, the spectra were contaminated by the emission from more extended sky regions. To take this into account, we first selected sky regions shown in grey shades on the background of Extended Data Fig.~\ref{fig:binning2}b.
The sky regions covered sufficiently large areas with a radius of $\sim 5$ arcmin around each detector region. The outer extent for the outermost region was slightly smaller, $\sim 3$ arcmin.
We then used the \texttt{xaarfgen} tool to perform ray-tracing simulations, which accounted for the shape of the Resolve mirror's PSF.

While using large sky regions for generating ARFs and SSM, it is important to emphasize that the gas properties measured from the spectra still largely reflect the properties of gas over smaller regions around the detector regions.
By running Monte-Carlo simulations with the \texttt{SIXTE} software package \cite{Dauser2019}, adopting the Hitomi/SXS PSF as an approximation for XRISM/Resolve, we modeled the regions emitting $90\%$ of photons that reach each detector region, indicated by white contours in Extended Data Fig.~\ref{fig:binning2}b.
These regions extend azimuthally up to $\simeq1-2\arcmin$ away from the detectors and are radially elongated due to the steep gradient of X-ray surface brightness.

To correct for the SSM when performing spectral modeling, we considered each spectrum $S_{\rm i}$ as a linear combination of contributions from all regions:
\be
S_i = R_i \left(\sum_{j=1}^{N_{\rm reg}}{A_{ij}M_{{\rm icm,}\,j}+A_{{\rm agn,}\,i}M_{\rm agn}}\right),
\ee
where $R_i$ is the RMF for the spectrum $S_i$, $A_{ij}$ and $A_{{\rm agn,}\,i}$ are ARFs for the detector region $i$ from each sky region $j$ or AGN, and $M_{{\rm icm,}\,j}$ and $M_{\rm agn}$ are the ICM and AGN models.
The details of this approach are described in \cite{Hitomi2018U}.
For simplicity, we set $A_{ij}=0$ for all on-sky  regions with contributions amounting to $<5\%$ of the total light in each respective detector region. Following this rule, we considered AGN contributions only in the two central detector regions (inner $\sim 40$ kpc of the cluster). For the default six-annuli binning scheme, for example, 33 ARFs were eventually adopted in our spectral fitting process.\\

\noindent\textbf{Spectral modeling of central AGN}\\
The bright and highly variable AGN within the central BCG, NGC~1275, is well-described by a power-law continuum, attenuated by galactic absorption, and a neutral Fe~K$\alpha$ fluorescent line at 6.4 keV \cite[see, e.g.,][]{Hitomi_agn, Reynolds_agn}.
Keeping the photon index of this power-law, $\Gamma$, free while fitting all the gas parameters, i.e., following the approach of \cite{Hitomi_agn}, gave high values of $\Gamma$, in the range of 2.1-2.3.
Based on the recently measured gamma-ray fluxes, the AGN is currently in non-flaring phase and the expected photon index is below 2 \cite{Fukazawa_2018}, which is in tension with the best-fitting value.
Additionally, such modeling with all parameters free resulted in high AGN fluxes, low normalization of the gas thermal component, and unphysically high ICM metallicities.

To overcome this issue, we constrained the AGN flux utilizing the non-variable ICM emission measured with Chandra.
We re-processed Chandra ObsIDs 3209, 4952, and 11714, following the standard procedure described by \cite{Vikhlinin2005}, and extracted spectra from the first three sky regions shown in Extended Data Fig.~\ref{fig:binning2}b, removing the inner $\sim 9$ arcsecond around NGC 1275 dominated by the AGN emission.
These spectra were modeled with the \texttt{TBabs*vapec} models in \texttt{Xspec} \cite{Arnaud1996} in the $4.0-7.3\keV$ energy range, fixing the HI column density at $1.38\times10^{21}\cm^{-2}$ and keeping all other abundances tied to Fe. We then constrained the AGN flux following two independent strategies.
\begin{itemize}

  \item In the first approach, we measured the ICM flux between $4-6$ keV - an energy range dominated by continuum emission - in each Chandra spectrum.
  We then modeled XRISM spectra from regions 1-3 by fixing the ratio of the ICM components' fluxes in the same $4-6$ keV range between regions 1/2 and 1/3 to the ratios measured with Chandra.
  The ICM component is otherwise modeled as described below.
  We left both the AGN slope and flux parameters free.
  Their best-fit 1, 2, and 3$\sigma$ contours are shown in Extended Data Fig.~\ref{fig:agn_params} (pink shaded regions).

 \item In the second approach, we first evaluated the equivalent width of Fe-He$\alpha$ complex ($W_{\rm eq}=F_{\rm line}/F_{\rm cont}$) in sky region 1 using Chandra data ($W_{\rm eq}=0.76\pm{0.01}$), where the $F_{\rm line}$ and $F_{\rm cont}$ are line and continuum fluxes between $6-7$ keV, respectively.
 We then used XRISM data to calculate the equivalent width in the same sky region by varying the AGN flux and photon index within the $26-36\times 10^{-12}\erg\s^{-1}\cm^{-2}$ and $1.4-2.0$ ranges, respectively, fixing both parameters each time.
 The blue contours in Extended Data Fig.~\ref{fig:agn_params} show the region in the $\Gamma$-flux space, where the XRISM ICM models provide an equivalent width consistent with Chandra's values within 1, 2, and 3$\sigma$.
\end{itemize}

The above two constraints are consistent at the $\sim3\sigma$ level within the parameter range $F_{2-10}\simeq30-32\times 10^{-12}\erg\s^{-1}\cm^{-2}$ and $\Gamma\simeq1.6-1.8$, which are similar to the values reported by Hitomi \cite{Hitomi_agn}.
For our subsequent analysis, we fixed the AGN flux to $F_{2-10\keV} = 31\times 10^{-12}\erg\s^{-1}\cm^{-2}$, guided by the experiments above, while allowing the photon index and fluorescent line width and redshift to vary.
The best-fit AGN photon index is $1.71^{+0.04}_{-0.05}$ (see the black point in Extended Data Fig.~\ref{fig:agn_params}), while the width and redshift of the fluorescent line are $19.2^{+5.8}_{-5.0}\eV$ and $1.750^{+0.094}_{-0.080}\times10^{-2}$, respectively.
We additionally confirmed that modifying the fixed AGN flux within a broader range ($28-34\times10^{-12}\erg\s^{-1}\cm^{-2}$) did not affect our measured gas velocities, which are largely insensitive to the AGN model.
\\

\noindent\textbf{Spectral modeling of thermal ICM}\\
The ICM emission from each sky region was modeled with a single-component emission spectrum from thermal plasma in collisional ionization equilibrium from AtomDB version 3.0.9 atomic database (\texttt{bvapec} model in \texttt{Xspec} version 12.14.1). Within the considered energy band (above 3 keV), the hot gas is well reproduced by such one-temperature model. He and C abundances were fixed at Solar values, the abundances of light elements from N to Si were tied to Fe, with all other parameters free. All abundances are relative to the \texttt{lpgs} proto-solar abundances \cite{Lodders2009}. To account for flux suppression due to resonant scattering \cite{Gilfanov1987,Hitomi_res_scat}, we excluded the Fe~He$\alpha$ resonance ($w$) line from the thermal component and modeled it separately with a redshifted Gaussian. In regions most affected by resonant scattering (regions 1-2), the width of the Gaussian component was left free to account for the line's heavier wings.
Finally, both the AGN and ICM components were corrected for the HI column density absorption along the LOS, with the $n_{\rm H}$ value fixed to $1.38\times10^{21}$ cm$^{-2}$ \cite{Absorption2013}. Extended Data Fig.~\ref{fig:specs_b2L} shows the best-fit models for each spectrum, including contributions from individual components, while Extended Data Table~\ref{tab:parameters} summarizes the best-fitting parameters and their uncertainties.
\\

\noindent\textbf{Spectral modeling of the non-X-ray background}\\
Non-X-ray background (NXB) spectra for each detector region were extracted from Resolve's night-Earth database using the \texttt{rslnxbgen} tool.
We applied the same screening criteria as for the on-source data and generated a diagonal RMF for the NXB spectra.
The typical NXB level is $< 10^{-3}$ $\rm counts\,\s^{-1}\keV^{-1}$ in the considered sub-FOV regions over the energy range used in the analysis (Extended Data Fig.~\ref{fig:specs_b2L}).

The NXB spectra for each region were modeled separately from the source spectra, using the diagonal RMF and no ARF.
We apply an empirical NXB model following \cite{Kilbourne2018}, consisting of a continuum (modeled by a single power-law) and 12 detector background emission lines in the 3-11 keV band (modeled by Gaussians).
We fit the continuum slope and normalization with all other parameters fixed, then fit the normalization of the emission lines.
The ratios of all emission lines other than Mn were fixed, as are the line widths, according to the recommended NXB model.
The best-fit model was then fixed and used as the NXB component while modeling the source spectra.
We confirmed that varying the overall normalization of the NXB by $\pm 50 \%$ does not alter our velocity measurements.\\

\noindent\textbf{Effects of Radial Binning}\\
To ensure the choice of a binning scheme does not bias our conclusion on radial velocity trends, we present two alternative radial binning methods, shown in the left panels of Extended Data Fig.~\ref{fig:vel_profs_check}.
In black, we show the radial profile of velocity dispersion and bulk velocity using the full FOV for each pointing. The coarser profiles do not show any significant deviations from the default velocity profiles, preserving the overall shape of the velocity structure.
We additionally show in blue the finer profile used in the velocity maps presented in Fig.~\ref{fig:map_bins} and Extended Data Fig.~\ref{fig:map_bins_error} (the regions along NW from the center).
While there are slight differences in the statistical uncertainties of each measurement, we confirm the overall amplitudes of velocities and the presence of gradients.
\\

\noindent\textbf{Systematic Uncertainties: XRISM Mirror Uncertainties} \\
In-flight calibration of Resolve and its X-ray Mirror Assembly is ongoing, thus there are still uncertainties in our understanding of spatial-spectral mixing in this analysis. Notably, the level of disagreement between ground calibration measurements and predictions from ray-tracing simulations can be significant, but it depends on the off-axis and azimuthal angles in ways that are not well-characterized. Based on our current information, we adopted a value of 30\% for the uncertainty in SSM. Given this systematic, we modeled the data while varying the normalization of the off-axis ARFs, and therefore the relative contribution of light from off-axis regions, by $\pm 30 \%$. For each additional fit, we repeated the AGN experiments described in the Methods above, finding that a $30 \%$ increase in SSM decreases the derived AGN flux by $10 \%$, and vice versa. We derived the corresponding velocity profiles presented in Extended Data Fig.~\ref{fig:vel_profs_check}b with notations equivalent to Fig.~\ref{fig:vel_profs}. With modified ARFs, each velocity measurement varies within the $1\sigma$ statistical uncertainties, not affecting any conclusions of our study.\\

\noindent\textbf{Spectroscopic Calibration Uncertainties}\\
The instrumental spectroscopic uncertainties that impact our velocity measurements are uncertainties in the broadband energy scale and the time-dependent gain reconstruction \cite{Eckart2024,Porter2016,Porter2024}, as well as in the energy resolution. These values have been assessed extensively on the ground and in-orbit. The energy scale uncertainty after gain reconstruction is estimated to be $0.4$ eV in the $5.4-8.0\keV$ band for these observations. This value is based on the $\sim 0.27$ eV gain reconstruction error at 5.9~keV and the systematic energy scale uncertainty of $\lesssim 0.3$~eV in this band, which are uncorrelated and can be root-sum-squared. The total uncertainty corresponds to $18\kms$ at $6.7\keV$, where the Fe-K$_{\alpha}$ lines contribute most of the constraint power. Over a broader energy range ($2-9\keV$), the total energy scale uncertainty is consistent with $\lesssim1$ eV.
The uncertainty on the per-pixel Hp energy resolution values provided in the CalDB is energy-dependent and estimated to be $\lesssim0.3$ eV FWHM from $2-9\keV$, which translates to $<10\kms$ uncertainty for the velocity dispersion entries in Extended Data Table~\ref{tab:parameters} \cite{Kitayama2014}. Its contribution to the systematic uncertainty of our measured velocity dispersion is smaller than the presented statistical error.
Finally, the XRISM observations were at most $7''$ offset from the nominally-requested pointings, with variations averaging around a few arcseconds.
\\

\noindent\textbf{Mach Numbers and Non-Thermal Pressure Support}\\
Using the measured gas temperature, $T$ (Extended Data Table~\ref{tab:parameters}), the sound speed of the gas was calculated as $c_{\rm s}=(\gamma k_{\rm B}T/\mu m_{\rm p})^{1/2}$, where $\gamma=5/3$ is the adiabatic index for monatomic gas, $k_{\rm B}$ is the Boltzmann constant, $\mu=0.61$ is the mean molecular weight in a fully ionized plasma, and $m_{\rm p}$ is the proton mass.
Combining this with the velocity dispersion, $\sigma_{\rm ICM}$, in each region, we determined the Mach number, $\mathcal{M}_{\rm\sigma}=\sqrt{3}\sigma_{\rm ICM}/c_{\rm s}$, changing between $\simeq 0.1$ and $0.3$, with the highest value reached in the innermost region.
Accounting for bulk velocities  \cite{Hitomi2018U}, $v_{\rm bulk}$, we found Mach numbers $\mathcal{M}_{\sigma\rm{+v}}=(3\sigma_{\rm ICM}^2+v_{\rm bulk}^2)^{1/2}/c_{\rm s}$ increased by less than 10-15\% in all regions except for region 3, where the increase was by $\simeq 30$\%.
These correspond to the kinetic to total (kinetic plus thermal) pressure ratio, $f_{\rm nth}=P_{\rm kin}/P_{\rm tot} = \mathcal{M}_{\rm\sigma}^2/(\mathcal{M}_{\rm\sigma}^2+3/\gamma)$, being $\simeq5\%$ in the innermost region and below $\simeq1\%$ in the third region with the lowest velocity dispersion (Extended Data Table~\ref{tab:parameters}).
\\

\noindent\textbf{X-ray Surface Brightness and Effective Length}\\
The X-ray emissivity of the Perseus cluster is peaked towards the center (Extended Data Fig.~\ref{fig:prof_leff}b, black curve).
Therefore, most of the flux (and spectrum) from a given sky region originates from an attenuated volume along the LOS rather than the entire LOS.
The LOS size of this finite region, or effective length $l_{\rm eff}$ \cite{Zhuravleva2012}, increases with the projected distance since X-ray emissivity rapidly decreases with radius (Extended Data Fig.~\ref{fig:prof_leff}a).
Taking the X-ray emissivity of Perseus and assuming spherical symmetry, we calculated $l_{\rm eff}$ as region size that contributes 50\% of the flux to the total flux at each projected distance from the cluster center (red curve in Extended Data Fig.~\ref{fig:prof_leff}b).
We also varied the contributing fraction by $\pm10\%$, showing the resulting effective lengths with a shaded region in the same figure. More conservatively, one can estimate the dominant length scale along the LOS as X-ray-emissivity-weighted length, $\int l\cdot\epsilon(l) dl$, where $\epsilon(l)$ is the normalized X-ray emissivity at a given distance $l$ along the LOS. Both estimates provide similar values for the dominant scale (Extended Data Fig.~\ref{fig:prof_leff}b).
\\

\noindent\textbf{Cooling and Heating Rates}\\
We estimated the dissipation heating rate as $Q_{\rm heat}=C_0\rho v_{\rm k}^3/\ell_k$ and the cooling rate as $Q_{\rm cool}=A_0\rho^2\Lambda(T)$ within the LOS range $[-\ell_{\rm eff}:\ell_{\rm eff}]$, both shown in Fig.~\ref{fig:heating_cooling_ratio}. Here, the gas mass density $\rho=(n_{\rm i}+n_{\rm e})\mu m_{\rm p}$, where $n_{\rm e}$ and $n_{\rm i}$ are the number densities of electrons and ions, respectively. The deprojected electron number density profile of Perseus was taken from the XMM-Newton measurements \cite{Churazov2003,Tang2017}. The number density of ions and electrons are related as $n_{\rm i}=(\xi-1)n_{\rm e}$, where $\xi=1.912$ for a fully ionized plasma. The normalized cooling function, $\Lambda(T)$, is fixed to $2.5\times10^{-23}\erg\cm^3\s^{-1}$ in this study \cite[for a solar metallicity gas;][]{Sutherland1993}. Its variations within the temperature range relevant to Perseus are small, less than $30$\% between the 3 and 7 keV temperatures. The coefficient in the cooling rate is $A_0=(\xi-1)/(\xi\mu m_{\rm p})^2$. Following \cite{Zhuravleva2014}, we adopted $C_0=3^{3/2}2\pi/(2C_{\rm K})^{3/2}\simeq5$, where $C_K\simeq 1.65$ is a fiducial number for the Kolmogorov constant \cite{Sreenivasan1995,Kaneda2003}. The specific constant value will depend on details of the velocity cascade, however, we are not concerned about its precise value as it is clearly an order-unity number.\\

\noindent\textbf{Numerical Simulations of Stratified Turbulence}\\
We performed 3D numerical simulations of hydrodynamical turbulence to support the interpretation of our velocity radial profiles, namely the inconsistency of the velocity dispersion radial profile with a single gas motion driver picture.
The simulations were carried out with the mesh-based MHD code FLASH4.6 \cite{Fryxell2000}.
In simulations, the ICM was assumed to initially be spherically symmetric and in hydrostatic equilibrium within a static gravitational potential.
Its initial density and temperature distributions followed the deprojected gas radial profiles from XMM-Newton observations \cite{Churazov2003,Tang2017}.
The simulation domain size was $600\kpc$ on each side, sufficiently large for our purpose of interpreting XRISM data.
During the simulations, turbulence was driven by a stochastic forcing term based on the Ornstein-Uhlenbeck process \cite{Eswaran1988,Schmidt2006,Federrath2010}, where we included only the solenoidal stirring mode dominant in galaxy clusters \cite[e.g.,][]{Miniati2015,Porter2015}.
Each of our simulations was run for at least $5\Gyr$ to allow turbulence to be fully developed within the simulated volume. The long-term turbulent heating slightly changes the gas density and temperature radial profiles of the ICM, which however does not affect any of our conclusions. Our default spatial resolution was $\simeq 2\kpc$ and two other resolutions $\simeq1$ and $4\kpc$ were tested to guarantee the convergence of our results.
In reality, ICM turbulence is driven by more complicated physical processes (e.g., multiple mergers, AGN feedback; see \cite{Simionescu2019} for a review). For Perseus, our simulations aimed to isolate the evolution of stratified random gas motions, given that the cluster had sufficient time to develop turbulence after the latest mergers (based on its nearly-relaxed dynamical state). More tailored simulations for modeling feedback- and sloshing-driven velocity spatial distributions in Perseus will be presented in our follow-up studies.

The single injection scale and energy amplitude were two parameters that determined the velocity power spectrum in our simulations.
We explored various injection scales: $\ell_{\rm inj}\simeq50$, $100$ and $500\kpc$, and found that $\ell_{\rm inj}\gtrsim100\kpc$ were required to capture the gradient of the velocity dispersion profile outside $\sim50\kpc$ (Extended Data Fig.~\ref{fig:sim_maps}), consistent with our expectations based on the isotropic turbulence cascade picture, even with gravitational stratification. For a (visually) direct comparison between simulations and observations, we chose the velocity amplitude parameter in the simulations to match the observed velocity dispersion near the radius $50-100\kpc$. In Extended Data Fig.~\ref{fig:sim_maps}a and \ref{fig:sim_maps}b, we show the simulated velocity dispersion distributions with $\ell_{\rm inj}\simeq500$ and $50\kpc$, respectively. Azimuthally elongated structures are observed in the former case, indicating the presence of stratified turbulence and gravity waves \cite[e.g.,][]{Brethouwer2007,Shi2019,Mohapatra2021}. By comparing the projected velocities with and without X-ray emissivity weight, we illustrated that the drop of X-ray velocity dispersion in the core is an observational effect, which has been utilized to probe velocities at different scales in our study.
\\

\noindent\textbf{Velocity radial profile of propagating sound waves}\\
Sound waves have long been proposed as a plausible heating mechanism associated with AGN feedback \cite[e.g.,][]{Fabian2003,Sanders2007}.
They are largely generated during the rapid expansion phase of X-ray bubbles and propagate outwards with the sound speed of $c_{\rm s}\simeq1000\kms$ in Perseus.
Considering a simple, spherically-symmetric situation without energy dissipation, following the model discussed in \cite{Fabian2017}, the energy flux of sound waves across the radius $r$ is conserved as $F_{\rm sw}=4\pi r^2\rho_{\rm gas}(r)v_{\rm s}(r)^2c_{\rm s}$, where $v_{\rm s}(r)$ is the wave velocity amplitude. We then estimated the X-ray-weighted velocity dispersion radial profile produced by these waves (Extended Data Fig.~\ref{fig:sound_waves}a), assuming that a fraction of the AGN energy deposits to the wave at the initial radius $r_0$ with a constant rate $F_{\rm sw}=\alpha F_{\rm agn}$, where $\alpha$ is a constant fraction and $F_{\rm agn}=5\times10^{44}\erg\s^{-1}$ is the AGN power \cite{Timmerman2022}. We fine-tuned $\alpha$ so that the modeled velocity dispersion within the innermost region (an X-ray emissivity–weighted average) matches our observational measurement.
For reasonable values of $r_0$ (essentially, the inner bubble radius), all modeled profiles are steeper than our measurements since the sound wave velocity amplitude attenuates rapidly. We note that, given the lifetime of the current innermost bubble pair ($\simeq10^8$ yr), the majority of sound waves within $r\lesssim100\kpc$ are driven by these bubbles, whose outer boundary is at $r\simeq15\kpc$ -- an upper limit of $r_0$.
If we further take dissipation into account, the modeled velocity radial velocity would become even steeper. We thus conclude that, based on our model, sound waves alone cannot explain the measured velocity dispersion radial profile within $\sim100\kpc$.

Meanwhile, whether and how sound waves dissipate efficiently if the ICM viscosity is much smaller than the Spitzer value are still open questions. We adopted an ad-hoc model to gain insights on this problem, while letting sound waves balance a fraction of cooling loss everywhere between the radius $r_0$ (fixing to $10\kpc$) and $r_{\rm max}$, i.e.,
\be
\frac{\dd F_{\rm sw}(r)}{\dd r}=\beta\cdot 4\pi r^2 Q_{\rm cool},
\ee
where $\beta$ is a constant fraction and $F_{\rm sw}(r_{\rm max})=0$.
Similar to the non-dissipation case, the predicted velocity dispersion profiles are much steeper than the XRISM measurements (Extended Data Fig.~\ref{fig:sound_waves}b).
The sound waves would induce a high velocity dispersion within $\sim20\kpc$ if they balance the cooling fully ($\beta=1$).
The measured $180\kms$ dispersion in our innermost region implies that no more than $\sim1/3$ of the cooling flux within $100\kpc$ is compensated by sound waves.
Such fraction is indeed likely smaller, since we assumed a highly fine-tuned heating model and the sound waves are not the only gas motion driver near the cluster center.
\\

\noindent
\textbf{Data availability}
The observational data analysed during this study will be available in the NASA HEASARC repository (https://heasarc.gsfc.nasa.gov/docs/xrism/) in the summer of 2025. The atomic databases used in this study are available online (AtomDB, http://www.atomdb.org/). \\

\noindent
\textbf{Code availability}
Publicly released versions of the FLASH Code are available via the Flash Center for Computational Science's website (https://flash.rochester.edu/).



\clearpage

\section*{}
\label{fig:ext}

\renewcommand{\figurename}{Extended Data Fig.}
\setcounter{figure}{0}
\renewcommand{\tablename}{Extended Data Table}
\setcounter{table}{0}

\begin{figure}
\centering
\includegraphics[width=0.95\linewidth]{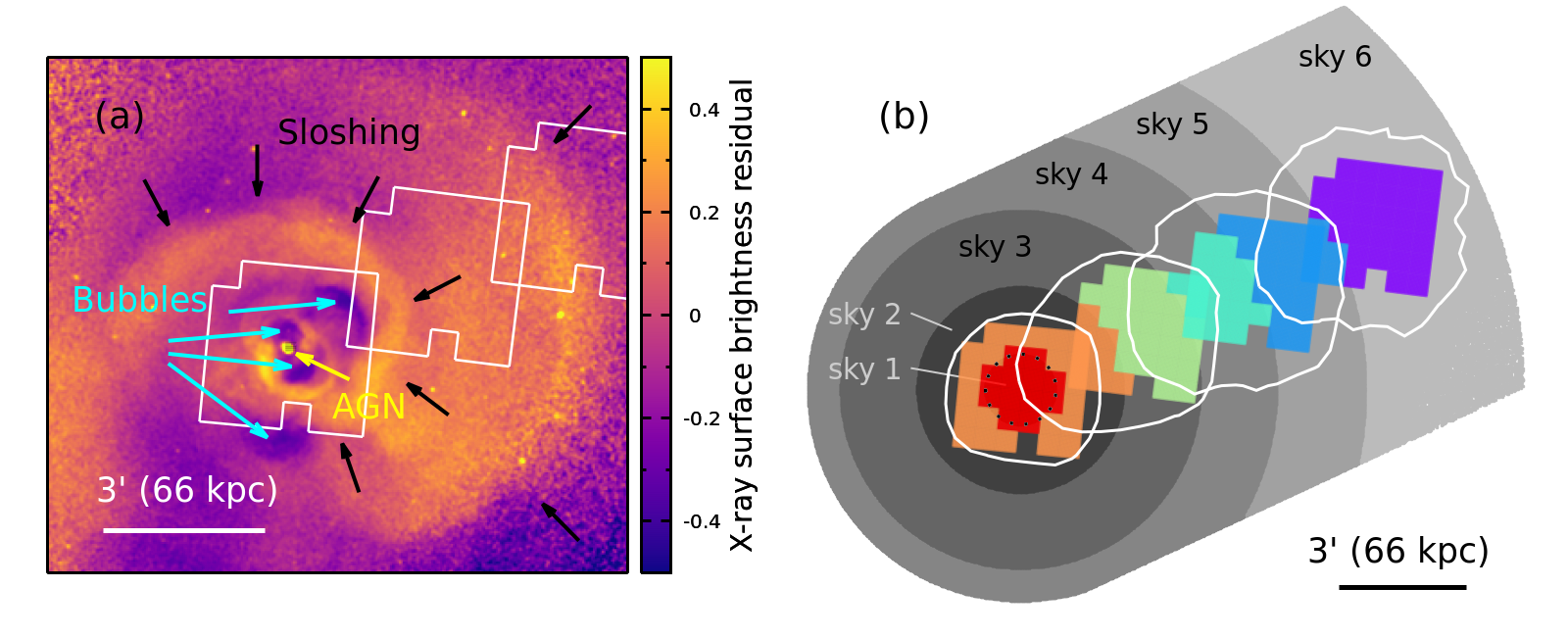}
\caption{\textbf{X-ray structures within the Perseus core and default radial binning scheme adopted for the analysis}. (a) Residual X-ray/Chandra image of the Perseus cluster in the $2-8\keV$ band, with arrows highlighting prominent structures relevant to our analysis. (b) The color pixels and grey circles/annuli show detector and sky regions (No. 1–6 from inner to outer), respectively, with the black dashed circle marking the innermost sky region. White contours illustrate the regions that contribute 90\% photons to each pointing based on our SIXTE simulations. }
\label{fig:binning2}
\end{figure}

\begin{figure}
\centering
\includegraphics[width=0.5\linewidth]{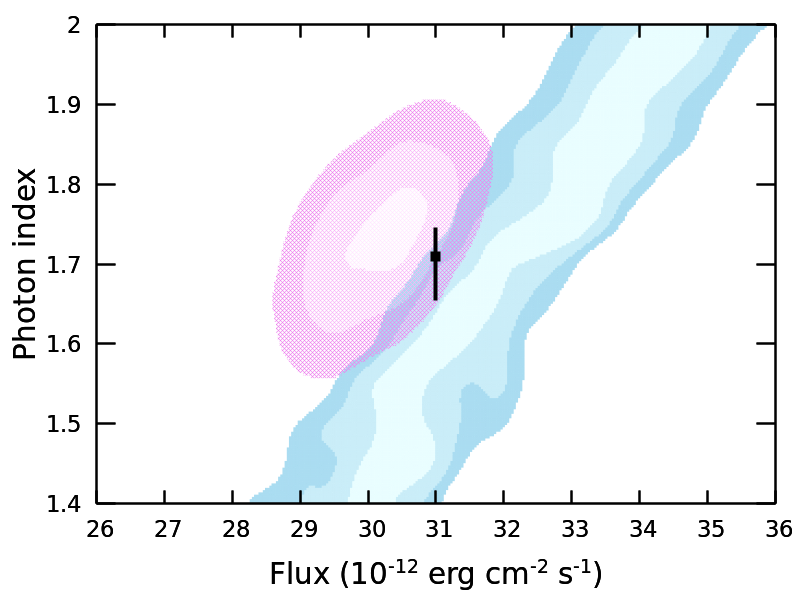}
\caption{\textbf{Constraints on AGN parameters: photon index $\Gamma$ vs. flux between $2-10\keV$. }
The contours indicate the implied 1, 2, $3\sigma$ parameter regimes from the two independent approaches utilizing Chandra spectra for the ICM component. Pink contours are based on the ratio of fluxes between sky regions 1/2 and 1/3 in the $4-6\keV$ band, which lacks strong emission lines, while blue contours are from the measurements of the equivalent width of Fe He-$\alpha$ line complex (Methods). For modeling the ICM in this work, we fixed the AGN flux at $31\times 10^{-12}\erg\s^{-1}\cm^{-2}$, shown as the black point with $1\sigma$ error bars of the photon index. }
\label{fig:agn_params}
\end{figure}

\clearpage

\begin{table*}
\centering
\renewcommand{\arraystretch}{1.5}
\begin{minipage}{\linewidth}
\centering
\caption{\textbf{Best-fit parameters of the ICM.} The data is modeled with the \texttt{TBabs*(bvapec+vzgauss)} model in the $3-11$ keV energy band. All errors reflect $1\sigma$ statistical uncertainties. The total cstat/dof = $152613/175930$. }
\label{tab:parameters}
\begin{tabular}{lcccccc}
  \hline
  Parameter & Reg 1 & Reg 2 & Reg 3 & Reg 4 & Reg 5 & Reg 6 \\
  \hline
Radii (arcmin) & 0 -- 0.9 & 0.9 -- 2.5 & 2.5 -- 4.3 & 4.3 -- 6.1 & 6.1 -- 8.2 & 8.2 -- 11.8 \\
  $k_{\rm B}T$ (keV)\footnote{The ICM temperature.} & $3.62^{+0.12}_{-0.11}$ & $3.98^{+0.05}_{-0.05}$ & $4.99^{+0.12}_{-0.06}$ & $5.75^{+0.21}_{-0.18}$ & $6.22^{+0.35}_{-0.27}$ & $7.82^{+0.48}_{-0.42}$ \\
  Fe abundance\footnote{In terms of the fraction of the solar value. The metal abundances of S, Ar, Ca, Fe, and Ni are free parameters in our model. The abundance of other than Fe elements will be reported in a follow-up study.}        &
   $0.65^{+0.03}_{-0.03}$ & $0.70^{+0.02}_{-0.01}$ & $0.50^{+0.03}_{-0.03}$ & $0.48^{+0.04}_{-0.04}$ & $0.39^{+0.05}_{-0.04}$ & $0.29^{+0.04}_{-0.05}$ \\
  z  ($\times10^{-2}$)\footnote{The ICM redshift.} & $1.788^{+0.004}_{-0.004}$ & $1.755^{+0.002}_{-0.002}$ & $1.708^{+0.003}_{-0.003}$ & $1.687^{+0.006}_{-0.006}$ & $1.726^{+0.007}_{-0.006}$ & $1.802^{+0.011}_{-0.010}$ \\
  $v_{\rm bulk}$ (km/s)\footnote{The bulk velocity of the ICM in the rest frame of the BCG.} & $152^{+10}_{-10}$ & $53^{+6}_{-6}$ & $-86^{+8}_{-8}$ & $-147^{+18}_{-17}$ & $-33^{+19}_{-18}$ & $197^{+32}_{-29}$ \\
  $\sigma_{\rm ICM}$ (km/s)\footnote{The ICM velocity dispersion.}       & $172^{+18}_{-16}$ & $135^{+9}_{-9}$ & $71^{+14}_{-15}$ & $186^{+20}_{-18}$ & $115^{+24}_{-25}$ & $188^{+32}_{-29}$ \\
  $\sigma_w$ (km/s)\footnote{The velocity dispersion measured from the Fe XXV He-$\alpha$ (W) line.}     & $189^{+18}_{-16}$ & $155^{+8}_{-12}$ & - & - & - & - \\
  $f_{\rm nth}$ ($\times10^{-2}$)\footnote{The non-thermal pressure fraction.}  & $5.0^{+1.0}_{-0.8}$ & $2.8^{+0.4}_{-0.3}$ & $0.6^{+0.3}_{-0.2}$ & $3.7^{+0.8}_{-0.7}$ & $1.3^{+0.6}_{-0.5}$ & $2.8^{+1.0}_{-0.8}$\\
  Norm ($\times10^{-2}$)\footnote{The normalization of the \texttt{bvapec} model.}  & $7.73^{+0.28}_{-0.28}$ & $19.8^{+0.33}_{-0.34}$ & $19.2^{+0.47}_{-0.44}$ & $7.64^{+0.22}_{-0.24}$ & $2.32^{+0.08}_{-0.10}$ & $1.90^{+0.06}_{-0.08}$ \\
\hline
\hline
\vspace{-18pt}
\end{tabular}
\end{minipage}
\end{table*}

\begin{figure*}
\centering
\includegraphics[width=0.95\linewidth]{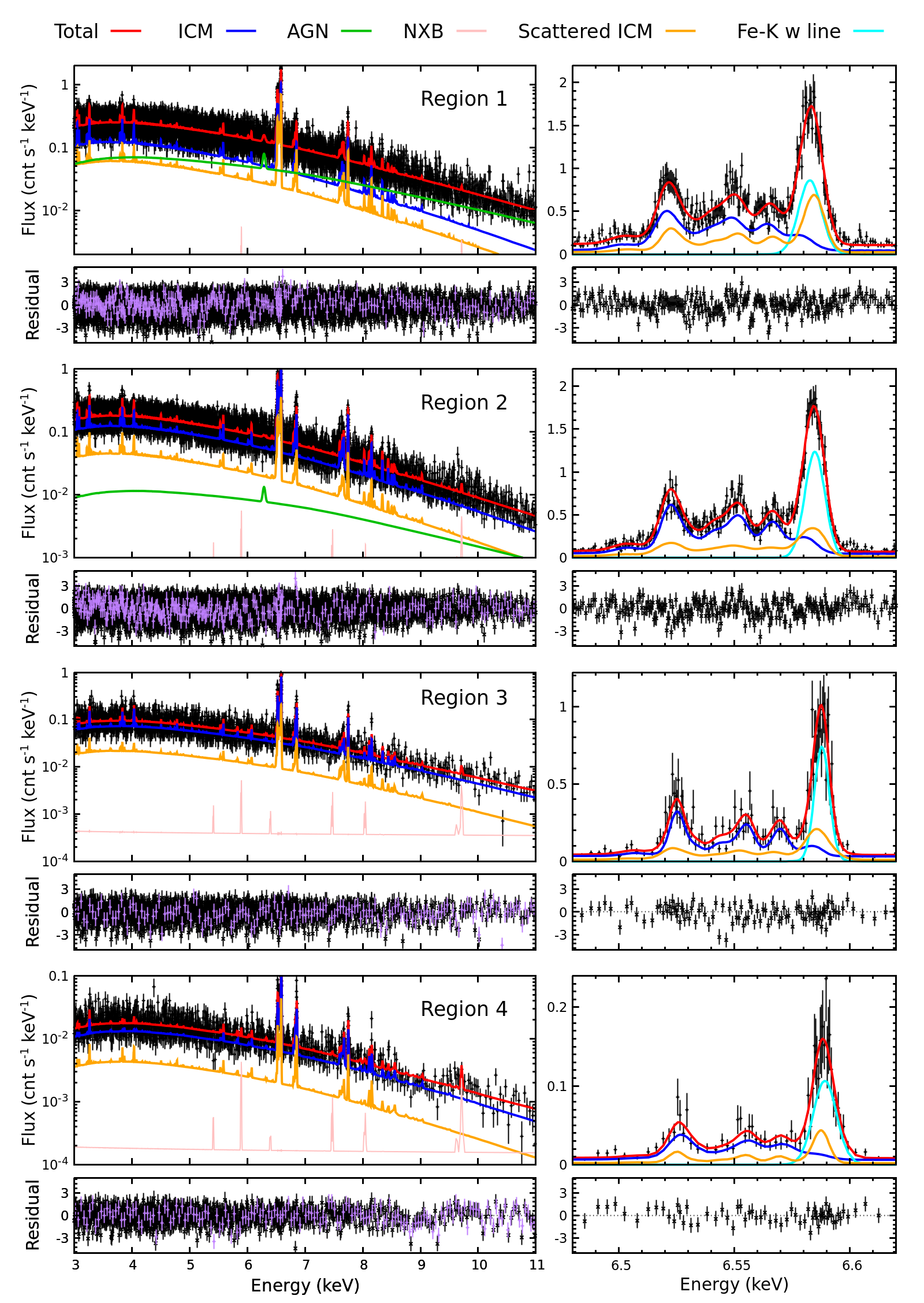}
%
\end{figure*}
\clearpage

\begin{figure*}
\centering
\includegraphics[width=0.95\linewidth]{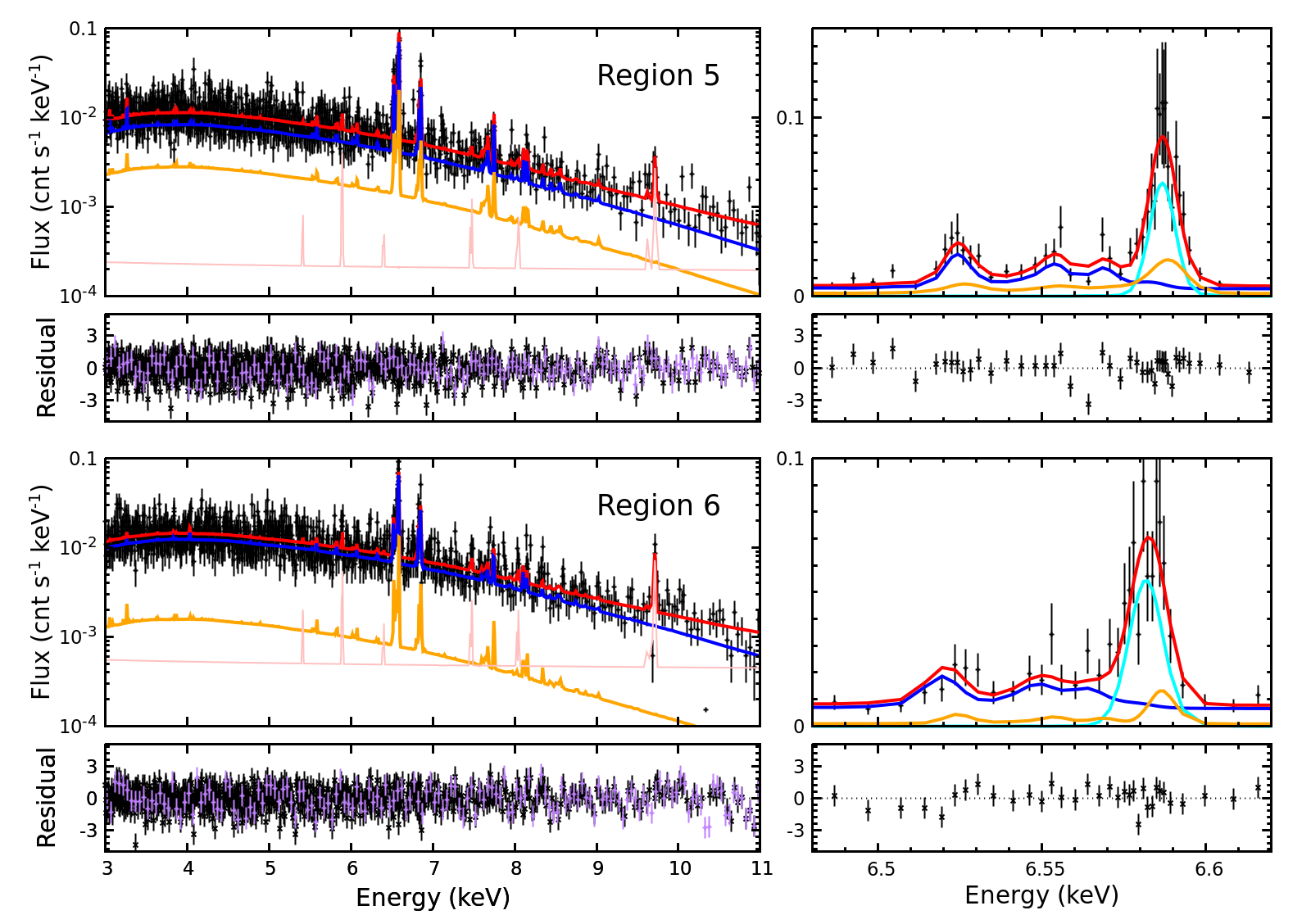}
\caption{\textbf{XRISM/Resolve spectra from each detector region and their best-fit models.} For each region, we show a broad-band spectrum from 3--11 keV and a detailed view of the Fe-K$_\alpha$ triplet on the right. Pink, orange, blue, and green curves represent the best-fit models of the NXB, the combined scattered ICM, the ICM, and the AGN components, respectively. The cyan curves on the right show the Fe-K $w$ line Gaussian component, while the red curves show the total models. Residuals normalized by the statistical errors, i.e., (data-model)/error, are displayed in the lower panels with two different binning schemes: black points correspond to the binning with minimum significance $\sigma_{\rm min}=3$, while purple points show the binning with $\sigma_{\rm min}=20$.
}
\label{fig:specs_b2L}
\end{figure*}

\begin{figure}
\centering
\includegraphics[width=0.95\linewidth]{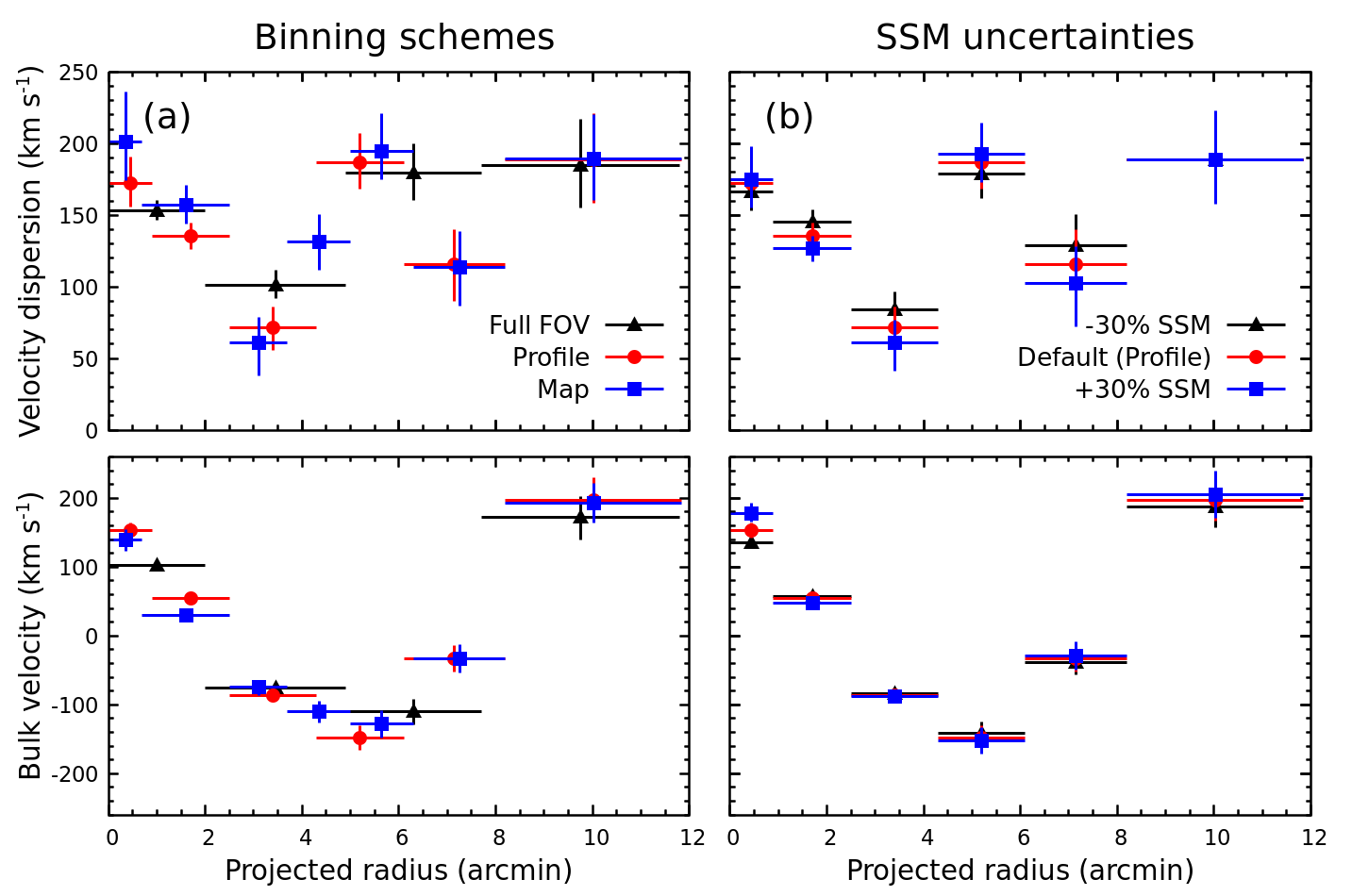}
\caption{\textbf{Gas radial velocity profiles for varied binning schemes (a) and spatial-spectral mixing (b). }
Formatting is equivalent to Fig.~\ref{fig:vel_profs}.
In panel (a), black triangles correspond to the full-FOV profile (broad radial bins), red circles to the nominal profile of Fig.~\ref{fig:vel_profs}, and blue squares to the profile of Fig.~\ref{fig:map_bins} (narrow radial bins).
In panel (b), the red circles are unchanged from Fig.~\ref{fig:vel_profs}, while blue squares/black triangles show profiles where off-axis ARFs, and therefore SSM, are increased/decreased by $30 \%$.}
\label{fig:vel_profs_check}
\end{figure}

\clearpage

\begin{figure}
\centering
\includegraphics[width=0.95\linewidth]{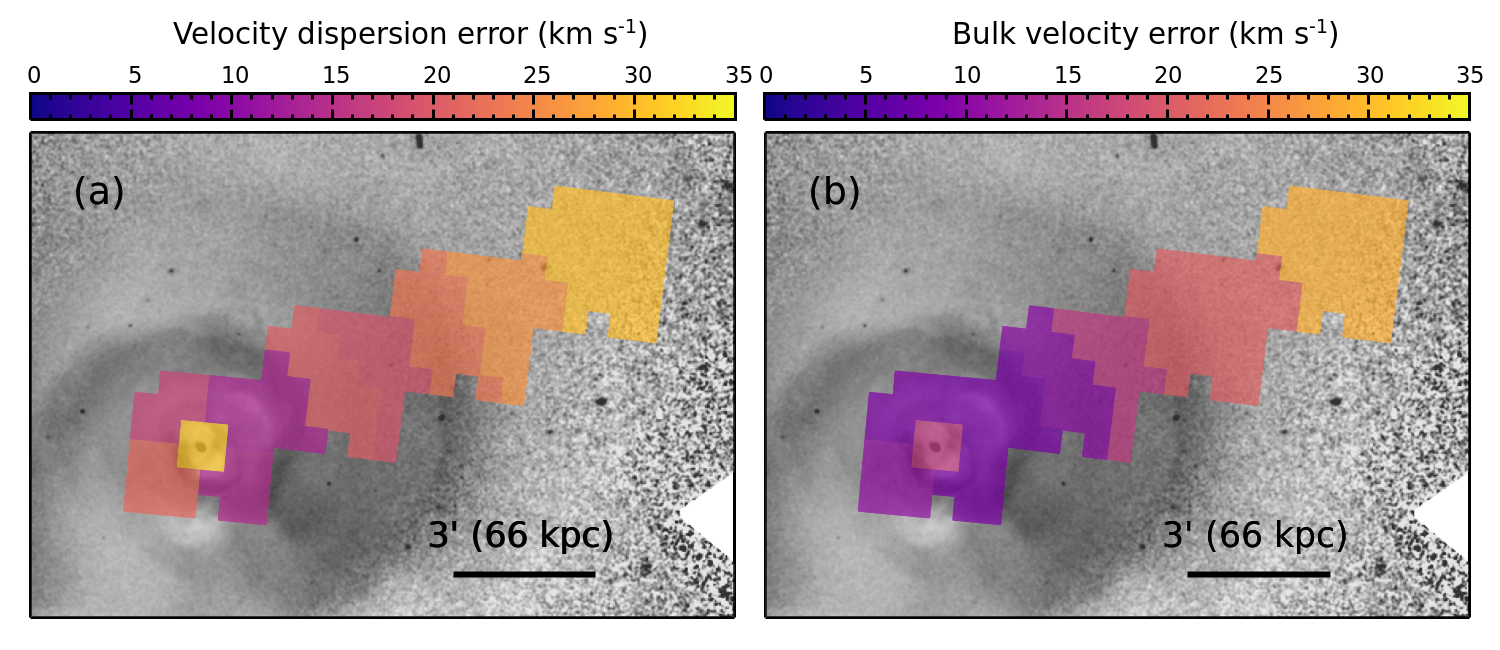}
\caption{\textbf{Statistical $1\sigma$ uncertainties for the velocity maps presented in Fig.~\ref{fig:map_bins}.} Velocity dispersion errors are shown on the left, while the bulk velocity errors are on the right. All other notations are the same as in Fig.~\ref{fig:map_bins}.}
\label{fig:map_bins_error}
\end{figure}

\begin{figure}
\centering
\includegraphics[width=0.95\linewidth]{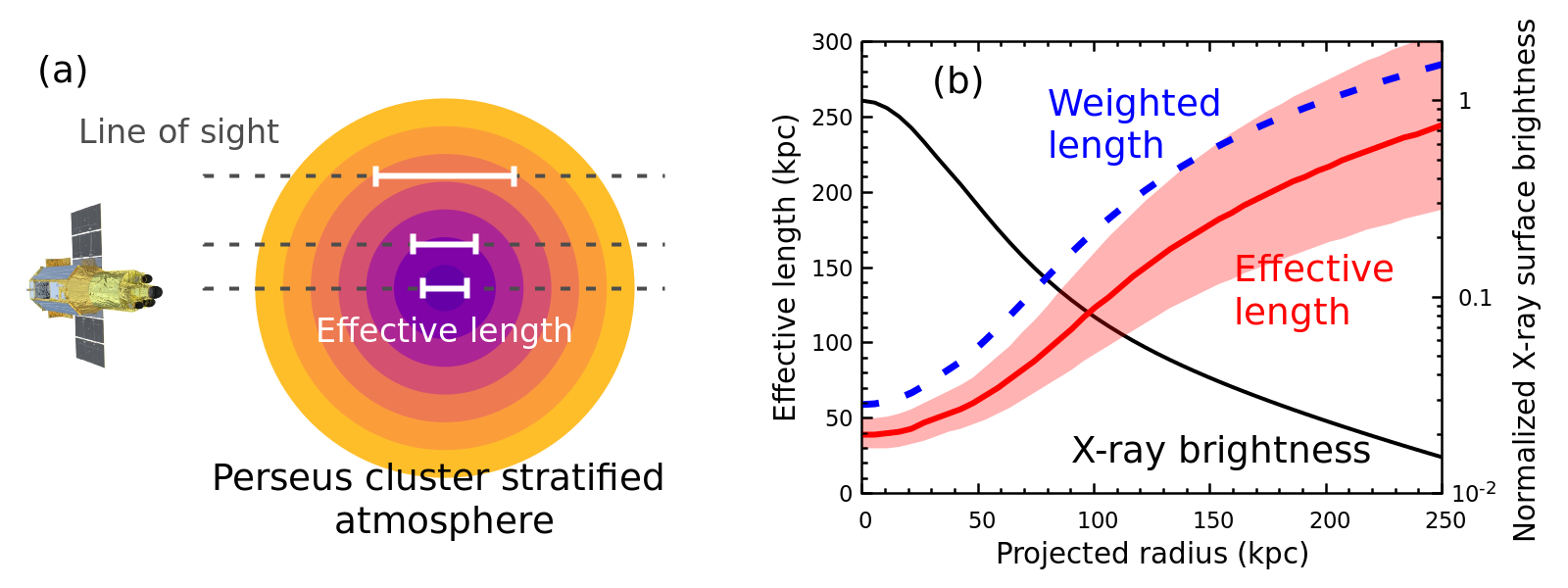}
\caption{\textbf{Effective length of the Perseus cluster.} (a) A sketch of the effective length concept in a stratified cluster atmosphere. The length along the LOS, which corresponds to the region size that contributes most to the measured flux/spectra, increases with the projected distance from the cluster center. (b) A radial profile of the normalized X-ray surface brightness in Perseus, adapted from \cite{Churazov2003}, is shown in black. The effective length, defined as the region size where 50\% of the flux is collected, is shown with the red curve. The shaded red region indicates the scales associated with 40-60\% of the flux contribution. The blue dashed curve represents the X-ray-weighted LOS length scale.}
\label{fig:prof_leff}
\end{figure}

\begin{figure}
\centering
\includegraphics[width=0.9\linewidth]{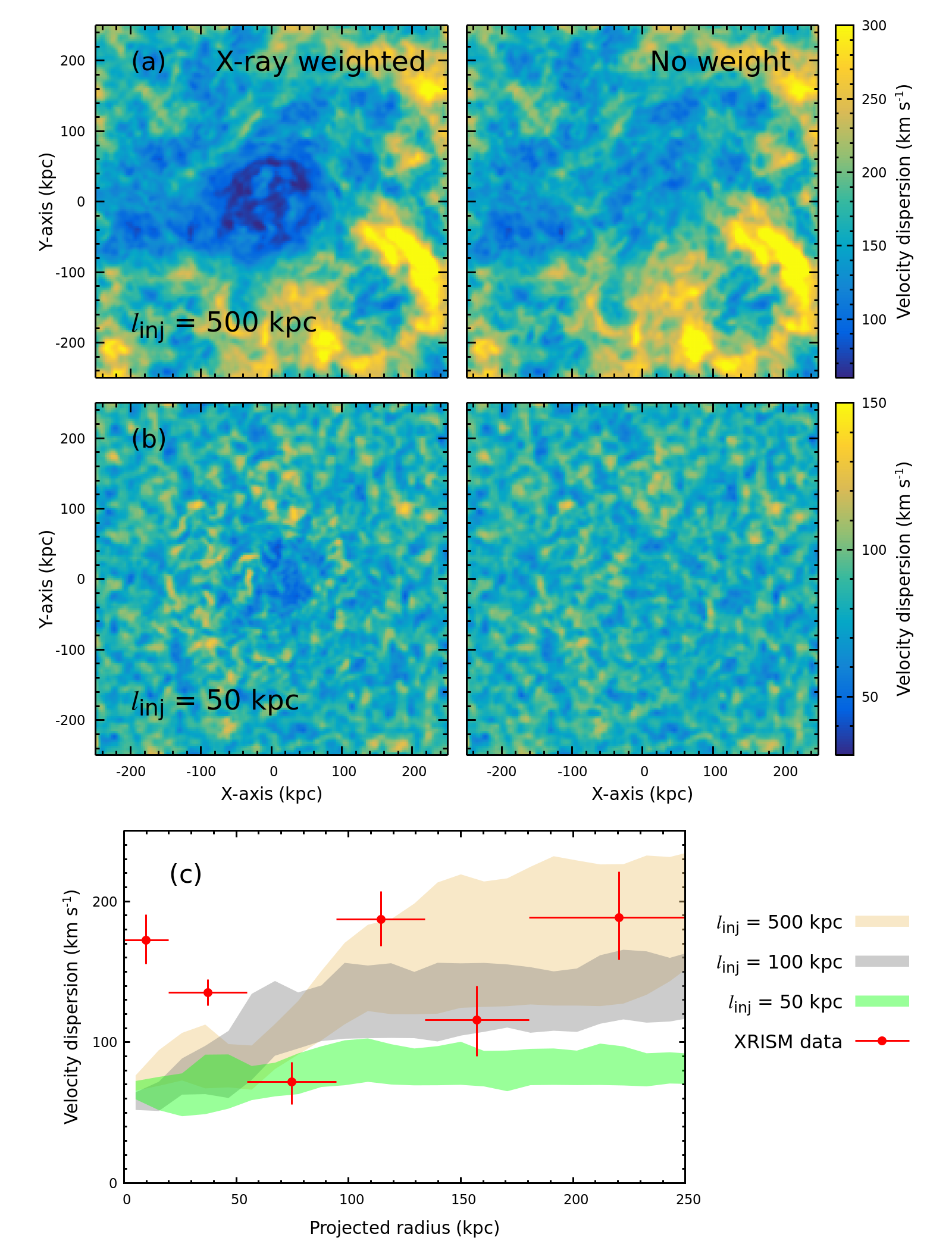}
\caption{\textbf{Numerical simulations of stratified turbulence with various driving scales.} (a) Projected velocity dispersion along the LOS for the simulation run with the injection scale $\ell_{\rm inj}=500\kpc$, weighted with (left) and without (right) X-ray emissivity, (b) Same as the panel (a) but for the run with $\ell_{\rm inj}=50\kpc$. (c) Comparisons between the simulated velocity dispersion radial profiles and the XRISM measurement (see also Fig.~\ref{fig:vel_profs}a). See Methods for the description of the simulations. }
\label{fig:sim_maps}
\end{figure}

\begin{figure}
\centering
\includegraphics[width=0.95\linewidth]{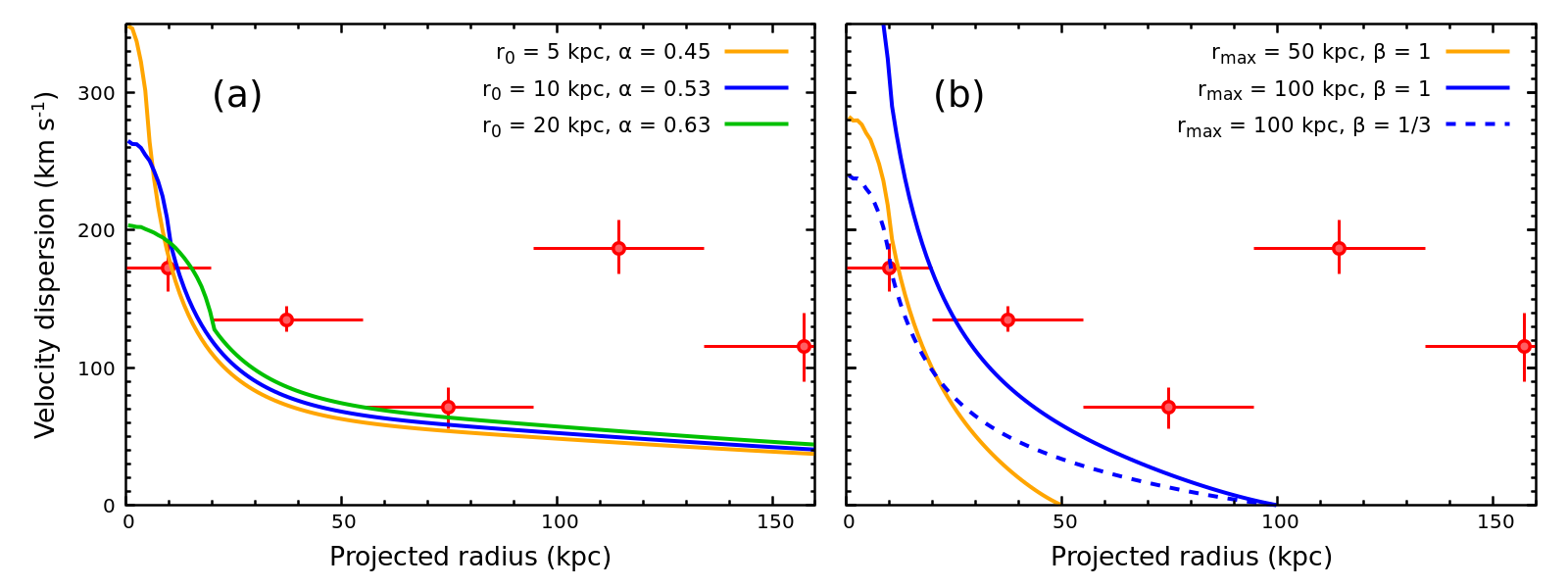}
\caption{\textbf{Modeled radial velocity dispersion profiles produced by propagating sound waves.} (a) Assuming conserved energy flux $F_{\rm sw}$ ($\equiv\alpha F_{\rm agn}$) of the sound waves starting from the cluster radius $r_0$ powered by the central AGN ($F_{\rm agn}=5\times10^{44}\erg\s^{-1}$). (b) Assuming sound waves compensate a fraction ($\beta$) of cooling loss everywhere within the radius $r_{\rm max}$. Red points are our measured velocity dispersion, same as in Fig.~\ref{fig:vel_profs}a. }
\label{fig:sound_waves}
\end{figure}

\clearpage

\noindent
\textbf{Acknowledgements}
%

This work was supported by JSPS KAKENHI grant numbers JP22H00158, JP22H01268, JP22K03624, JP23H04899, JP21K13963, JP24K00638, JP24K17105, JP21K13958, JP21H01095, JP23K20850, JP24H00253, JP21K03615, JP24K00677, JP20K14491, JP23H00151, JP19K21884, JP20H01947, JP20KK0071, JP23K20239, JP24K00672, JP24K17104, JP24K17093, JP20K04009, JP21H04493, JP20H01946, JP23K13154, JP19K14762, JP20H05857, and JP23K03459. 
Additional support came from NASA grant numbers 80NSSC23K0650, 80NSSC20K0733, 80NSSC18K0978, 80NSSC20K0883, 80NSSC20K0737, 80NSSC24K0678, 80NSSC18K1684, 80NNSC22K1922.
RB, RB-M, KH, TH, ML, FM, KM, AO, and KP acknowledge support from NASA under award number 80GSFC21M0002.
AB was supported by JSPS KAKENHI Grant Number JP23H01211.
EB acknowledges support from NASA grants 80NSSC24K1148 and 80NSSC24K1774.
LC acknowledges support from NASA grant 80NSSC25K7064.
CD acknowledges support from STFC through grant ST/T000244/1.
RF was supported by JSPS KAKENHI Grant Number JP23K03454.
LG acknowledges financial support from the Canadian Space Agency (grant 18XARMSTMA).
TR thanks the Waterloo Centre for Astrophysics and generous funding to B.R.M. from the Canadian Space Agency and the National Science and Engineering Research Council of Canada.
YM was supported by JSPS KAKENHI grant number JP23K22548.
MM acknowledges support from Yamada Science Foundation.
PP acknowledges support from NASA grants 80NSSC18K0988 and 80NSSC23K1656 and NASA contract NAS8-0360.
MS acknowledges the support by the RIKEN Pioneering Project Evolution of Matter in the Universe (r-EMU) and Rikkyo University Special Fund for Research (Rikkyo SFR).
AT and the present research are in part supported by the Kagoshima University postdoctoral research program (KU-DREAM).
YT was supported by the Strategic Research Center of Saitama University.
SY acknowledges support by the RIKEN SPDR Program.
IZ, AH, and CZ acknowledge partial support from the Alfred P. Sloan Foundation through the Sloan Research Fellowship. IZ performed part of the work at the Kavli Institute for Theoretical Physics (KITP) supported by grant NSF PHY-2309135.
JHL acknowledges the Canadian Space Agency (CSA) grant 22EXPXRISM.
SU acknowledges the support from the National Science and Technology Council of Taiwan (111-2112-M-001-026-MY3).
CZ was supported by the GACR grant 21-13491X.
TY and AT acknowledge support by NASA under award number 80GSFC24M0006.

Part of this work was performed under the auspices of the U.S. Department of Energy by Lawrence Livermore National Laboratory under Contract DE-AC52-07NA27344. 
This work was supported by the JSPS Core-to-Core Program, JPJSCCA20220002. 

\\

\noindent
\textbf{Author Contributions}
Congyao Zhang, Annie Heinrich, and Irina Zhuravleva performed data analysis, explored systematic uncertainties, interpreted the results, and prepared the manuscript. Congyao Zhang also developed supporting numerical simulations. As the leader of the Perseus cluster target team in the XRISM Science Team, Irina Zhuravleva oversaw the work on the project. F. Scott Porter and Megan Eckart calibrated the energy scale and gain and estimated calibration uncertainties. Anna Ogorzalek, Francois Mernier, Shutaro Ueda, Julian Meunier, Yuto Ichinohe contributed to the Resolve data analysis. Maxim Markevitch, Anna Ogorzalek,  Kotaro Fukushima, Shogo Kobayashi, Kyoko Matsushita, Richard Mushotzky, Yasushi Fukazawa contributed to resolving the AGN contribution in spectral modeling. Caroline Kilbourne assisted with the NXB modeling and contributed to related discussions. Tahir Yaqoob provided valuable insights into the XRISM mirror uncertainties. Benjamin Vigneron and Julie Hlavacek-Larrondo provided the SITELLE analysis of the multiphase gas. Elena Bellomi, Megan Eckart, Yutaka Fujita, Julie Hlavacek-Larrondo, Yuto Ichinohe, Maxim Markevitch, Kyoko Matsushita, Brian McNamara, Francois Mernier, Anna Ogorzalek, Naomi Ota, F. Scott Porter, Aurora Simionescu, Phillip C Stancil, Nhut Truong, Shutaro Ueda, Benjamin Vigneron, John ZuHone reviewed the manuscript and contributed to discussions. The scientific goals of XRISM were discussed and developed over seven years by the XRISM Science Team, all members of which are authors of this paper. All the instruments were prepared by the joint efforts of the team. \\

\noindent
\textbf{Author Information}
Reprints and permissions information is available at www.nature.com/reprints. The authors declare no competing financial interests.
Correspondence and requests for materials should be addressed to Congyao Zhang (cyzhang@astro.uchicago.edu), Annie Heinrich (amheinrich@uchicago.edu) and Irina Zhuravleva (zhuravleva@uchicago.edu). \\

\clearpage

\noindent
\textbf{XRISM Collaboration}
\vspace{3mm}

\noindent
XRISM Collaboration$^{1}$,
Marc Audard$^{2}$,
Hisamitsu Awaki$^{3}$,
Ralf Ballhausen$^{4,5,6}$,
Aya Bamba$^{7}$,
Ehud Behar$^{8}$,
Rozenn Boissay-Malaquin$^{9,5,6}$,
Laura Brenneman$^{10}$,
Gregory V.\ Brown$^{11}$,
Lia Corrales$^{12}$
% ...

\vspace{3mm}
\noindent
$^1$Corresponding Authors: Congyao Zhang (cyzhang@astro.uchicago.edu), Annie Heinrich (amheinrich@uchicago.edu) and Irina Zhuravleva (zhuravleva@uchicago.edu).
$^2$Department of Astronomy, University of Geneva, Versoix CH-1290, Switzerland,
$^3$Department of Physics, Ehime University, Ehime 790-8577, Japan,
$^4$Department of Astronomy, University of Maryland, College Park, MD 20742, USA,
$^5$NASA / Goddard Space Flight Center, Greenbelt, MD 20771, USA,
$^6$Center for Research and Exploration in Space Science and Technology, NASA / GSFC (CRESST II), Greenbelt, MD 20771, USA,
$^7$Department of Physics, University of Tokyo, Tokyo 113-0033, Japan,
$^8$Department of Physics, Technion, Technion City, Haifa 3200003, Israel,
$^9$Center for Space Science and Technology, University of Maryland, Baltimore County (UMBC), Baltimore, MD 21250, USA,
$^{10}$Center for Astrophysics | Harvard-Smithsonian, MA 02138, USA,
$^{11}$Lawrence Livermore National Laboratory, CA 94550, USA,
$^{12}$Department of Astronomy, University of Michigan, MI 48109, USA,
% ...

\end{document}